\documentclass[aps,nofootinbib,preprintnumbers]{revtex4-1}

\usepackage{amssymb,amsmath}
\usepackage{graphicx}

\newcommand{\be}{\begin{equation}}
\newcommand{\ee}{\end{equation}}
\newcommand{\as}{\alpha_{\mathrm{s}}}
\newcommand{\Qs}{Q_{\mathrm{s}}}
\newcommand{\gev}{\;\mathrm{GeV}}
\newcommand{\tev}{\;\mathrm{TeV}}

\newcommand{\bo}[1]{\boldsymbol{#1}}
\newcommand{\bp}{\boldsymbol{p}}
\newcommand{\bk}{\boldsymbol{k}}
\newcommand{\bl}{\boldsymbol{l}}
\newcommand{\bw}{\boldsymbol{w}}
\newcommand{\bq}{\boldsymbol{q}}
\newcommand{\pperp}{\boldsymbol{p}_\perp}
\newcommand{\kperp}{\boldsymbol{k}_\perp}

\newcommand{\Kperp}[1]{\boldsymbol{k}_{#1 \perp}}

\newcommand{\lperp}{\boldsymbol{l}_\perp}
\newcommand{\qperp}{\boldsymbol{q}_\perp}
\newcommand{\wperp}{\boldsymbol{w}_\perp}
\newcommand{\pperpsq}{\boldsymbol{p}^2_\perp}
\newcommand{\kperpsq}{\boldsymbol{k}^2_\perp}

\newcommand{\LO}{{\rm{LO}}}
\newcommand{\LLog}{{\rm{LLog}}}

\newcommand{\rhot}{\tilde{\rho}}

\begin{document}

\preprint{INT-PUB-14-022}

\title{Triple- and Quadruple-Gluon Azimuthal Correlations from Glasma and Higher-Dimensional Ridges}

\author{\c{S}ener \"{O}z\"{o}nder}
\email{ozonder@uw.edu}
\affiliation{Institute for Nuclear Theory, University of Washington, Seattle, WA 98195, USA}

\date{\today}

%\pacs{
%25.75.-q,  % heavy-ion nuclear reactions - relativistic
%25.75.Gz, % Particle correlations, relativistic collisions, 
%12.38.Mh, % Plasmas quark-gluon
%25.75.Ld  % Collective flow, relativistic collisions
%}

%%%%
%%%%
%%%%
\begin{abstract}
We calculate the triple- and quadruple-gluon inclusive distributions with arbitrary rapidity and azimuthal angle dependences in the gluon saturation regime by using glasma diagrams. Also, we predict higher-dimensional ridges
in triple- and quadruple-hadron correlations for p--p and p--Pb collisions at LHC, which have yet to be measured.
In p--p and p--Pb collisions at the top LHC energies, gluon saturation is expected to occur since smaller Bjorken-$x$ values are being probed. 
Glasma diagrams, 
which are enhanced at small-$x$, include the gluon saturation effects, and they are used for calculating the long-range rapidity correlations (``ridges'') and $v_n$ moments of the azimuthal distribution of detected hadrons. The glasma description reproduces the systematics of the data on both p--p and p--Pb ridges.
As an alternative,
 relativistic hydrodynamics has also been applied to these small systems quite successfully. With the triple- and quadruple-gluon azimuthal correlations, this work aims to set the stage by going beyond the double-gluon azimuthal correlations in order to settle unambiguously the origin of ``collectivity'' in p--p and p--Pb collisions. We derive the triple- and quadruple-gluon azimuthal correlation functions in terms of unintegrated gluon distributions at arbitrary rapidities and azimuthal angles of the produced gluons. Then, unintegrated gluon distributions from the running coupling Balitsky-Kovchegov evolution equation are used to calculate the triple- and quadruple-gluon correlations for various parameters of gluon momenta, initial scale for small-$x$ evolution and beam energy.

\end{abstract}

\maketitle

%%%%%
%%%%%
%%%%%
\section{Introduction}

In nucleus-nucleus collisions, 
ridges occur in
the di-hadron correlations 
as structures
that are elongated in pseudorapidity difference $\Delta \eta = \eta_1 - \eta_2$, and 
peak
on the near $\Delta \phi \sim 0$ and away sides $\Delta \phi \sim \pi$,
where $\Delta \phi_1 - \phi_2$ is the azimuthal angle difference between hadron pairs.
The origin of these correlations
have long been ascribed to the interplay between these two effects: 
The rapidity correlations between the gluons that are produced from the same longitudinal color flux tube
and the radially outward collective flow due to quark gluon plasma giving rise to the azimuthal collimation and anti-collimation of hadron pairs in the near and away sides, respectively. Historically, the appearance of collective flow as the prime indicative of quark gluon plasma have been thought to be specific to nucleus-nucleus collisions, and ridges had not been observed experimentally, nor had they been predicted by event generators particularly for p--p collisions 
until recently.

This established idea \cite{Dumitru:2008wn} has been challenged when the CMS collaboration at LHC announced the discovery of the ridge in high-multiplicity p--p collisions at ${\sqrt{s}=7\tev}$ \cite{Khachatryan:2010gv,Velicanu:2011zz,Li:2012hc}. Soon after, this was followed by the discovery of the ridge in p--Pb collisions at ${\sqrt{s}=5.02\tev}$ \cite{CMS:2012qk,Abelev:2012ola,Aad:2012gla,Milano:2014qua,ABELEV:2013wsa,Abelev:2014mda,Abelev:2014mva}. This puzzling situation gave rise to the question whether
quark gluon plasma has been created even in small systems such as p--p and p--Pb. Relativistic hydrodynamics that employ the collective flow idea has been applied to these small systems, and interesting results have been obtained for p--Pb \cite{Bozek:2011if,CMS:2012qk,Bozek:2012gr,Bozek:2013ska,Werner:2013ipa,Kozlov:2014fqa,Bzdak:2014dia} and p--p collisions \cite{Bozek:2010pb}.

Another approach to explain the ridges in p--p and p--Pb collisions comes from saturation physics, and particularly its implementation within the glasma framework. Glasma (``glassy plasma'') refers to the flux tubes of classical $SU(3)$ color fields which result from high density color charge in the projectile and target at small-$x$. High gluon density in a nucleon or nucleus also gives rise to the emergence of the semi-hard saturation scale $\Qs$, which itself makes diagrammatic approach feasible. Recently, the double-gluon glasma diagrams have been combined with the double-gluon BFKL diagrams, which are important at small-$x$ regardless of the gluon saturation. The variation of the ridge signal in the data for various momentum windows and multiplicity classes (``systematics'') in p--p and p--Pb collisions has been reproduced successfully without resorting to
any collective flow in hydrodynamical sense \cite{Dumitru:2010iy,Dusling:2012iga,Dusling:2012cg,Dusling:2012wy,Dusling:2013oia,Venugopalan:2013cga}.

All these recent developments invite further studies that can possibly distinguish the underlying mechanism of the ridges. By deriving the triple- and quadruple-gluon azimuthal correlation functions in the gluon saturation regime we aim to take the first step towards this direction. For quantitative predictions on the triple- or quadruple-hadron spectra, one must convolve these gluon correlation functions with fragmentation functions, which will be performed in a separate study. Triple- and quadruple-hadron spectra can also be computed from hydrodynamics simulations, and data on these higher order hadron correlations can be another ground for testing the two frameworks; glasma and hydrodynamics. 

Currently both glasma and hydrodynamics are seen as competing approaches (see the discussion in Ref. \cite{Venugopalan2014}). However, even in the case that hydrodynamics as a coarse grained model captures the liquid behavior of possible quark gluon plasma in p--p and p--Pb collisions, it would be desired to know how collectivity arises on a more fundamental level. Since successful applications of hydrodynamics to water do not preclude the existence of more fundamental van der Waals forces, we can ask what the van der Waals forces of  quark gluon plasma might be.

In perturbative QCD, hadronic and nuclear collisions are understood as interaction of two partons from the projectile and target, which themselves are modeled in terms of parton distribution functions. One then adds radiative corrections to this picture by resumming logarithms of virtuality $Q^2$, Bjorken-$x$ or both simultaneously. Depending on the kinematic regime of the events, one of the resummation schemes such as DGLAP, BFKL, Double Logarithmic Approximation is followed. 
However, the conventional parton distribution functions constructed this way do not include gluon saturation.
The two main effects of gluon saturation are slowdown of the unbounded growing of the gluon density, and the non-kinematical, nontrivial correlations between produced gluons depending on their transverse momenta and rapidities. In this work, we use unintegrated gluon distributions (UGD) from the running coupling Balitsky-Kovchegov (rcBK) evolution equation; these UGDs include gluon saturation effects. The non-kinematical correlation effect is captured by  glasma diagrams which we shall calculate to triple- and quadrupole-gluon order. Any final quantitative study should combine gluon productions from both perturbative QCD and glasma diagrams.

This paper is organized in a way that first the well-known single- and double-gluon azimuthal cumulants are reviewed. By this, we aim to clarify the confusion in the literature regarding the prefactors, and in the mean time we will be showing the steps of glasma calculations over the relatively easier cases of the single- and double-gluon inclusive distributions. Then, we move to the derivation of the triple- and quadruple-gluon inclusive distributions. Finally we show the results of our numerical calculations where we use the UGDs obtained from the rcBK equation. We leave the quantitative predictions on the triple- and quadruple-hadron azimuthal correlations for another study; this requires convolving the triple- and quadruple-gluon correlations with  fragmentation functions to attain final hadron spectra. However, 
our results based on gluons and how they change
with the variance of transverse momentum and rapidity as well as the number of participants as presented in the last section can be seen as qualitative predictions for the future measurement on triple- and quadrupole-hadron correlations at LHC for p--p and p--Pb collisions.

%%%%%
%%%%%
%%%%%
\section{Review of the Single- and Double-Gluon Azimuthal Cumulants from Glasma}

In this section we first give an overview of the single- and double-gluon inclusive distribution functions from glasma. The purpose of this section is introducing the main steps of diagram calculation in glasma as well as 
determining the correct
prefactors of the azimuthal cumulants which is mostly unsettled in the literature (See Appendix \ref{appxPrefactors}).

\begin{figure}[t]
\begin{centering}
$\vcenter{\hbox{\includegraphics[scale=0.60]{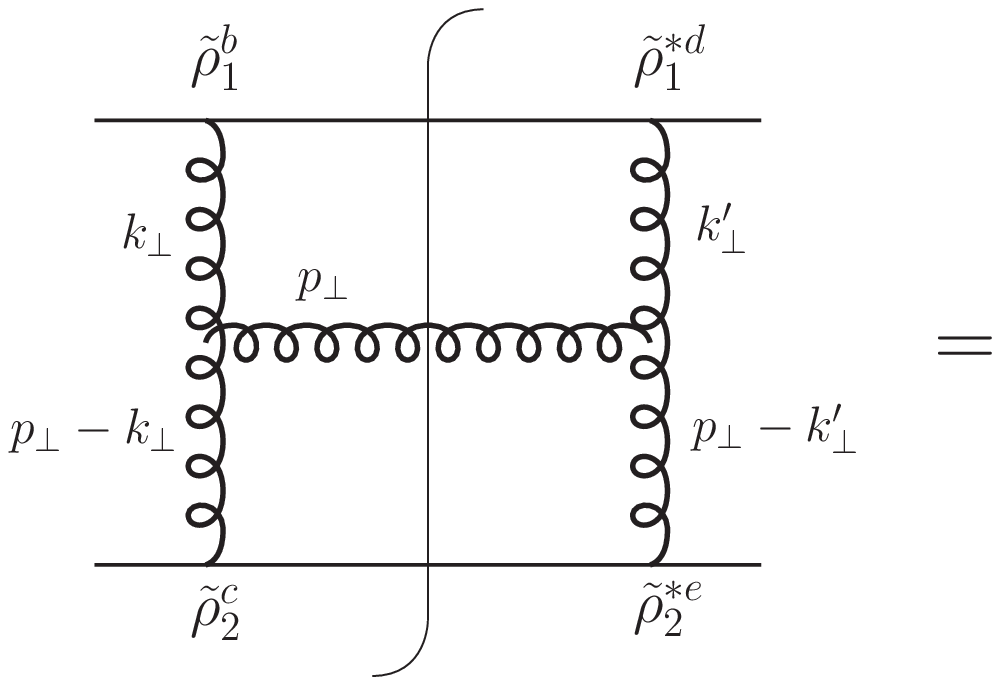}}}$
$\vcenter{\hbox{\includegraphics[scale=0.60]{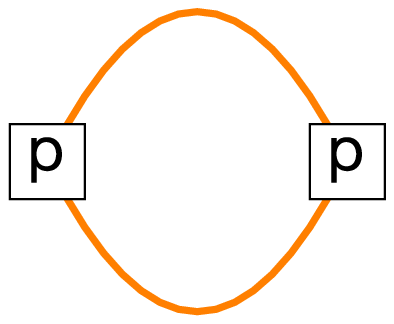}}}$
\par\end{centering}
\caption{(left) The diagram for the single gluon production from the classical field $A^a_\mu (p)$ created by the color charge densities of the target $\tilde{\rho}_1$ and projectile $\tilde{\rho}_2$. (right) The dimerized form of the same diagram on the left (the cut line is not shown). The upper connecting curve is for the charge correlations in the target and the lower one is that for the projectile. }
\label{fig:single}
\end{figure}

We start with the single-gluon inclusive  distribution function. The rate for a single gluon in the leading order  is given by \cite{Blaizot:2004wu}
\be
N_g \Big|_\LO = \int \frac{d^3 \bo{p}}{(2\pi)^3 2 E_p} \sum_{a,\lambda}  | {\cal M}^a_\lambda (\bo{p}) |^2,
\label{N1}
\ee
where the gluon production amplitude is
\be
{\cal M}^a_\lambda (\bo{p}) = p^2 A^{a,\mu}(p) \epsilon_\mu^{(\lambda)}(\bo{p}). 
\label{amp1}
\ee
Here $p=(E,\bo{p})$, $A^{a,\mu}(p)$ is the classical gluon field with color index $a$, and it satisfies the classical Yang-Mills equation with the sources of color charge density $\rho^a_1(\bo{x}_\perp)$ and $\rho^a_2(\bo{x}^\prime_\perp)$ of the projectile and target.
The polarization vector of the produced gluon with polarization $\lambda$ is denoted by $\epsilon_\mu^{(\lambda)}$. 
By transforming the Lorentz invariant phase space factor as
${d^3 \bo{p}/ E_p \longrightarrow  d^2 \bo{p}_\perp d y_p}$,
one can write the leading order single-gluon inclusive distribution as
\be
\frac{d N_1[\tilde{\rho}^a_{1,2}]}{d^2 \bo{p}_\perp d y_p} \Bigg|_\LO = \frac{1}{2 (2 \pi)^3} \sum_{a,\lambda}   | {\cal M}^a_\lambda (\bo{p}) |^2.
\label{dN1}
\ee 
Classical gauge fields due to the color charge density of the target and projectile in the leading order in powers of $ \tilde{\rho}^a_{1,2} / \bo{k}^2_\perp$ can be written as follows\footnote{Equation~(\ref{Aexpn}) appeared with an extra factor of $1/2$ in Ref.~\cite{Dumitru:2008wn}; however, this mistake has been corrected in a footnote in 
Ref.~\cite{Gelis:2009wh}.} \cite{Dusling:2009ni,Blaizot:2004wu,Kovner:1995ts,Kovchegov:1997ke}
\be
p^2 A^{a,\mu}(p) = -i f^{abc} g^3 \int \frac{d^2 \bo{k}_\perp}{(2 \pi)^2} L^\mu(\bp,\kperp)\frac{\tilde{\rho}^b_1(\kperp) \tilde{\rho}^c_2(\pperp - \kperp)}{\kperpsq (\pperp - \kperp)^2}, 
\label{Aexpn}
\ee
where $f^{abc}$ is the structure constant of the gauge group, $g$ is the gauge coupling and $L^\mu$ is the Lipatov vertex whose components in the light-cone coordinates are given by 
\begin{align}
L^+(\bp,\kperp)= & -\kperp^2/p^-, \\ 
L^-(\bp,\kperp)= & \left[(\pperp-\kperp)^2-\pperpsq \right]/p^+, \\
L^i(\bp,\kperp)= & -2 \kperp^i. 
\end{align}
By substituting Eq.~(\ref{Aexpn}) into Eq.~(\ref{dN1}) and making the replacement $\sum_\lambda \epsilon_\mu^{* (\lambda)} \epsilon_\nu^{(\lambda)}\rightarrow -g_{\mu \nu}$, the single-gluon inclusive distribution can be written as 
\begin{align}
\frac{d N_1}{d^2 \bo{p}_\perp d y_p} \Bigg|_\LO =  -\frac{1}{2 (2 \pi)^3}  & g^6 f_{abc} f_{ade} 
\int \frac{d^2 \kperp}{(2 \pi)^2} \frac{d^2 \kperp^\prime}{(2 \pi)^2} L^\mu(\bp,\kperp) L_\mu(\bp,\kperp^\prime) \nonumber \\
& \times \frac{\tilde{\rho}^b_1(\kperp) \tilde{\rho}^c_2(\pperp - \kperp)}{\kperpsq (\pperp - \kperp)^2}
\frac{\tilde{\rho}^{*d}_1(\kperp^\prime) \tilde{\rho}^{*e}_2(\pperp - \kperp^\prime)}{\bo{k}_\perp^{\prime 2} (\pperp - 
\kperp^\prime)^2}.
\label{dN1expn}
\end{align}
The glasma diagram corresponding to the single gluon production is shown in Fig.~\ref{fig:single}.

In a given collision $\tilde{\rho}^a_1$ and $\tilde{\rho}^a_2$ are not known; so for any quantity one proceeds by taking the average of it over the ensemble containing all possible color charge configurations of the projectile and target,
\be
\left< {\cal O}  \right>_{\LLog }= \int [D\rho_1] [D\rho_2] W[\rho_1] W[\rho_2] {\cal O}[\rho_1,\rho_2].
\ee
Here the weight function $W[\rho_{1,2}]$ evolves with JIMWLK renormalization group equation \cite{JalilianMarian:1996xn,JalilianMarian:1997jx,JalilianMarian:1997gr,JalilianMarian:1998cb,Iancu:2000hn,Iancu:2001ad,Ferreiro:2001qy} and it already includes resummations in powers of $\as \ln(1/x_{1,2})$, where $x_{1,2}$ are the Bjorken-$x$ values of the partons from the projectile and target \cite{Dusling:2009ni,Gelis:2008rw,Gelis:2008ad,Gelis:2008sz}.
In the approximation of the local (white noise) Gaussian fluctuations of the McLerran-Venugopalan model  \cite{Dumitru:2008wn,McLerran:1993ni,*McLerran:1993ka,*McLerran:1994vd,Kovchegov:1996ty,Ozonder:2012vw}, the average of multiplication of many  charge densities as in Eq.~(\ref{dN1}) can be written in terms of the two-point correlation function
\be
%\begin{align}
%\langle \tilde{\rho}^a(\kperp) \rangle & = 0, \\
\langle \tilde{\rho}^a(\kperp) \tilde{\rho}^{*b}(\kperp^\prime) \rangle_{A_1,A_2} = (2\pi)^2 \mu^2_{A_1,A_2}(y_p,\kperp)\delta^{ab} \delta^2 (\kperp - \kperp^\prime), \label{corr1} 
%\end{align}
\ee
where $\mu^2(y_p,\kperp)$ is the Fourier transform of the color charge squared per unit transverse area, and $y_p$ is the momentum rapidity of the produced gluon. Here only the densities  from the same nucleon or nucleus are correlated. After averaging Eq.~(\ref{dN1expn}) over all possible color charge configurations, and by using Eq.~(\ref{corr1}) and ${f_{abc}f_{abc}=N_c (N_c^2-1)}$, one arrives at
\be
\left< \frac{d N_1}{d^2 \bp_\perp d y_p}  \right>_\LLog = \frac{- g^6 S_\perp N_c (N_c^2-1)}{2 (2\pi)^3}
\int \frac{d^2 \kperp}{(2\pi)^2} \frac{L^\mu (p,\kperp) L_\mu (p,\kperp)}{\kperp^4 (\pperp-\kperp)^4} \mu^2_{A_1}(y_p,\kperp) \mu^2_{A_2}(y_p,\pperp - \kperp),
\label{dN1b}
\ee
where $S_\perp$ is the transverse area of the overlap between the target and projectile, and it results from the replacement
\be
\int d^2 \kperp \delta^2(\kperp - \kperp^\prime) \delta^2(\kperp - \kperp^\prime) \longrightarrow  \frac{S_\perp}{(2\pi)^2} \int d^2 \kperp \delta^2(\kperp - \kperp^\prime).
\ee
The unintegrated gluon distribution per unit transverse area (UGD) is defined as \cite{Dusling:2009ni}
\be
\Phi_{A_{1,2}} (y_p,\pperp) \equiv g^2 \pi (N_c^2-1)\frac{\mu^2_{A_{1,2}}(y_p,\pperp)}{\pperp^2}.
\label{ugd}
\ee
By using Eq.~(\ref{ugd}) and $L^\mu (p,\kperp) L_\mu (p,\kperp)=-4 \kperp^2 (\pperp-\kperp)^2 / \pperp^2$, Eq.~(\ref{dN1b}) becomes\footnote{Our result here matches with the one in Ref. \cite{Dusling:2009ni}.}
\be
\left< \frac{d N_1}{d^2 \bp_\perp d y_p}  \right>_\LLog = \frac{\as  N_c S_\perp}{\pi^4  (N_c^2-1) }
\frac{1}{\pperp^2} \int \frac{d^2 \kperp}{(2\pi)^2}  \Phi_{1,p}(\kperp) \Phi_{2,p}(\pperp - \kperp),
\label{dN1result}
\ee
where we used the compact notation $\Phi_{1,p}(\kperp) \equiv \Phi_{A_1}(y_p,\kperp)$.

\begin{figure}[t]
\begin{centering}
$\vcenter{\hbox{\includegraphics[scale=0.50]{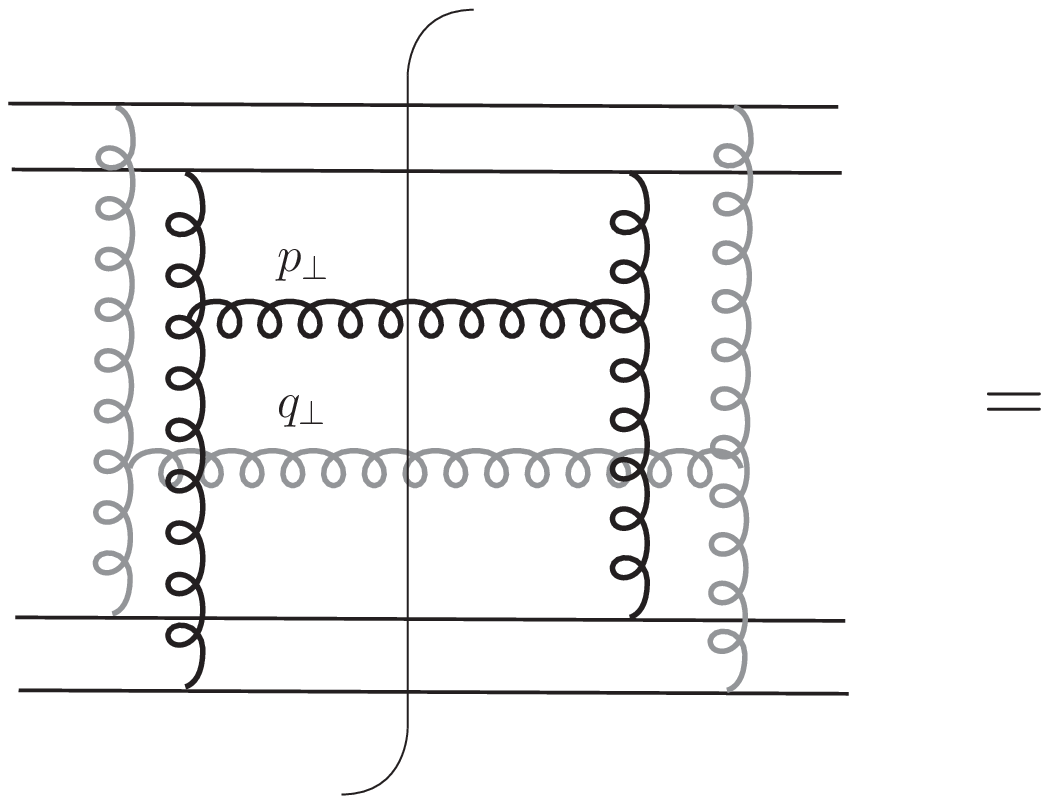}}}$
$\vcenter{\hbox{\includegraphics[scale=0.80]{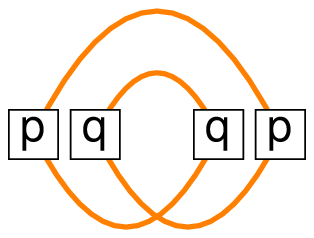}}}$\\
$\vcenter{\hbox{\includegraphics[scale=0.50]{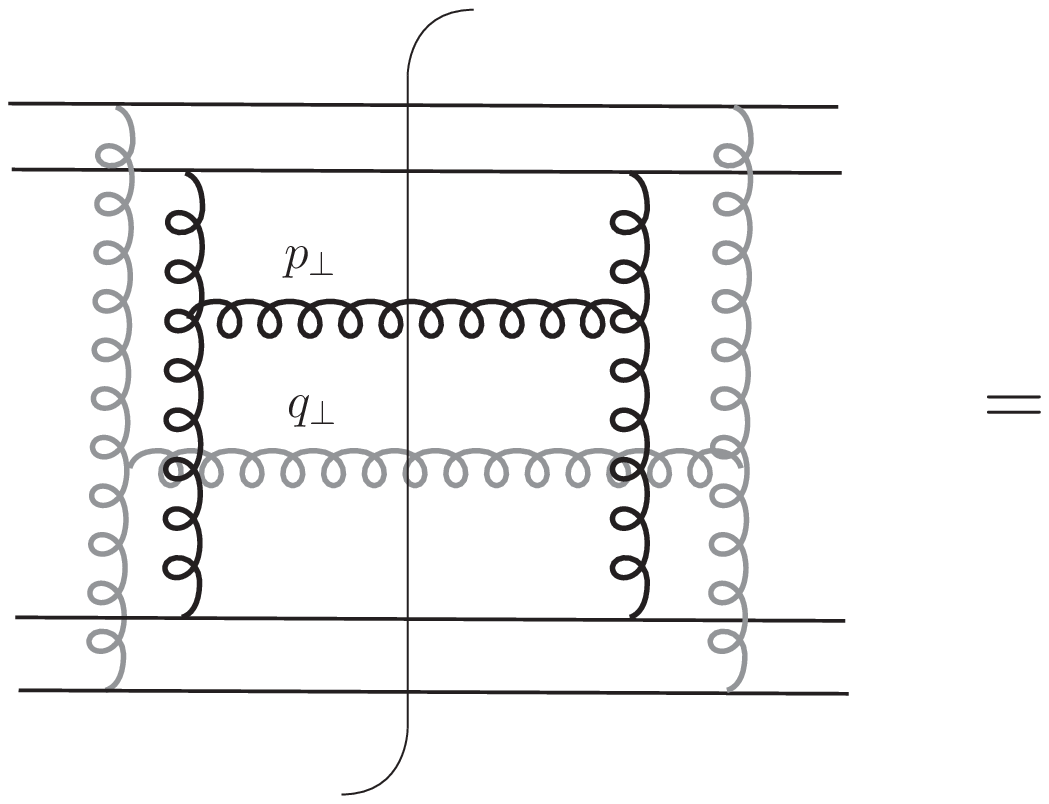}}}$
$\vcenter{\hbox{\includegraphics[scale=0.63]{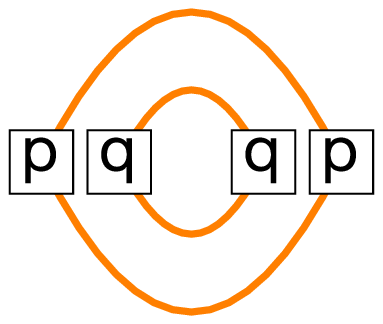}}}$
\par\end{centering}
\caption{(up) One of the eight connected glasma diagrams for double gluon production in the conventional and dimerized forms. (down) The only disconnected diagram, which is contained in $\left<  d N_2 \middle/ d^2 \bo{p}_\perp d y_p d^2 \bq_\perp d y_q \right>$, but subtracted in $C_2(\bp,\bq)$.}
\label{fig:double}
\end{figure}

We now turn to the calculation of the double-gluon inclusive distribution from classical color fields in the leading logarithmic order. We start with
\be
\left<  \frac{d N_2}{d^2 \bo{p}_\perp d y_p d^2 \bq_\perp d y_q} \right>_\LLog  = \frac{1}{2^2 (2 \pi)^6} \sum_{a,a^\prime, \lambda, \lambda^\prime} \left<  | {\cal M}^{a a^\prime}_{\lambda \lambda^\prime} (\bp,\bq) |^2  \right>,
\label{dN2}
\ee 
where
\be
{\cal M}^{a a^\prime}_{\lambda \lambda^\prime} (\bo{p},\bo{q}) = p^2 q^2 A^{a,\mu}(p) A^{a^\prime,\sigma}(q) \epsilon_\mu^{(\lambda)}(\bo{p}) \epsilon_\sigma^{(\lambda^\prime)}(\bo{q}). 
\label{amp2}
\ee
After using Eq.~(\ref{Aexpn}), we end up with
\be
\Big< \rhot_1^b(\bk_{1 \perp})  \rhot_1^d(\bk_{3 \perp}) \rhot_1^{*h}(\bk_{4 \perp}) \rhot_1^{*f}(\bk_{2 \perp})  \Big>
 \Big< \rhot_2^c(\pperp - \bk_{1 \perp})  \rhot_2^e(\qperp - \bk_{3 \perp})  \rhot_2^{*i}(\qperp -\bk_{4 \perp}) \rhot_2^{*g}(\pperp - \bk_{2 \perp}) \Big>.
 \label{double-gluon-contractions}
\ee
This leads to 9 distinct diagrams in total as there are three different contractions for each of the projectile and target.
In Fig.~\ref{fig:double} we show one of the 8 connected diagrams along with the disconnected diagram. 
Out of the 8 connected diagrams, we shall only consider the ``rainbow'' diagrams  as in Ref.
\cite{Dusling:2009ni}; there are 4 such diagrams.  Rainbow diagrams are the leading diagrams where dimers with the same momentum index are contracted among each other either on the upper or lower part of a given diagram, as in the connected diagram in Fig.~\ref{fig:double}. The tools for contractions and diagram calculation are explained 
in Appendix \ref{appxDiagrams}.

The disconnected diagram in Fig.~\ref{fig:double} can be seen as two separate single-gluon diagrams as shown in Fig.~\ref{fig:single}. A more useful quantity is the connected double-gluon correlation function which is defined as 
\be
C_2(\bp,\bq) \equiv \left< \frac{d N_2}{d^2 \bo{p}_\perp d y_p d^2 \bq_\perp d y_q} \right> -  
\left< \frac{d N_1}{d^2 \bo{p}_\perp y_p} \right> \left< \frac{d N_1}{d^2 \bo{q}_\perp y_q} \right>,
\label{C2}
\ee
where the subscript ``LLog'' will be suppressed henceforth. 
The disconnected contribution that is contained in the first term in Eq.~(\ref{C2}), which contains 4 connected and 1 disconnected diagram, is subtracted by the second term.
Hence calculating only the 4 connected diagrams will directly yield second cumulant $C_2(\bp,\bq)$.

In the calculation of $C_2(\bp,\bq)$ one should consider not only the contractions between the dimers
across the cut line, but also contractions among the dimers on the same side of the cut line. Hence, all of the correlators below are needed
\begin{align}
\langle \tilde{\rho}^a(\kperp) \tilde{\rho}^{*b}(\kperp^\prime) \rangle_{A_1,A_2} & =  (2\pi)^2 \mu^2_{A_1,A_2}(y_{p,q},\kperp)\delta^{ab} \delta^2 (\kperp - \kperp^\prime) \label{corr1again}, \\
\langle \tilde{\rho}^a(\kperp) \tilde{\rho}^{b}(\kperp^\prime) \rangle_{A_1,A_2} &= (2\pi)^2 \mu^2_{A_1,A_2}(y_{p,q},\kperp)\delta^{ab} \delta^2 (\kperp + \kperp^\prime), \label{corr2}  \\
\langle \tilde{\rho}^{*a}(\kperp) \tilde{\rho}^{*b}(\kperp^\prime) \rangle_{A_1,A_2} &= (2\pi)^2 \mu^2_{A_1,A_2}(y_{p,q},\kperp)\delta^{ab} \delta^2 (\kperp + \kperp^\prime). \label{corr3}
\end{align}
We use 
 Eq.~(\ref{corr1again}) when two dimers from opposite sides are contracted, and Eqs.~(\ref{corr2}) and (\ref{corr3}) when two dimers from the same side are contracted. The rapidity index of $\mu^2$ in Eqs.~(\ref{corr1again}), (\ref{corr2}) and (\ref{corr3}) will be determined as follows. We choose a convention that $y_p$ is closer to the rapidity of the projectile and $y_q$ is closer to the rapidity of the target. Hence, when contracting two charge densities, one connected to the gluon with momentum $\bp$ and the other to the gluon with momentum $\bq$ (see Fig.~\ref{fig:double}), one should take $y_p$ if the charge densities have nucleus index $1$ (projectile), and $y_q$ is the they have nucleus index $2$ (target)
\begin{align}
\left< \rhot(y_p) \rhot(y_q) \right>_{A_1}  \propto \mu^2_{A_1}(y_p),\\
\left< \rhot(y_p) \rhot(y_q) \right>_{A_2}  \propto \mu^2_{A_2}(y_q). \label{rapidity-structure}
\end{align}
This convention is independent of whether the $\tilde{\rho}$'s are the starred ones or not.

The rest of the calculation is similar to that of the single gluon except that here we also use $f^{abc} f^{abd}=N_c \delta^{cd}$. The double-gluon inclusive distribution have the form\footnote{The mistake in the prefactor in 
Ref.~\cite{Dusling:2009ni} has been corrected in Ref.~\cite{Dusling:2012iga} where ``$\int_{\kperp}$'' means ``$\int d^2 \kperp$'' without a factor of $1/(2 \pi)^2$.
As for Eq.~(12) in Ref. \cite{Gelis:2009wh}, it does not reproduce correctly neither the prefactor of Eq.~(\ref{C_2}) here nor the momentum dependence of the UGDs in Eq.~(\ref{DA12}).}
\be
C_2(\bp,\bq)  = \frac{\as^2  N_c^2 S_\perp}{\pi^{8}  (N_c^2-1)^3}
\frac{1}{\pperp^2 \qperp^2} \int \frac{d^2 \kperp}{(2\pi)^2}  (D_1+D_2),
\label{C_2}
\ee
where
\begin{align}
D_1 & = \Phi_{1,p}^2 (\kperp) \Phi_{2,p}(\pperp - \kperp) D_{A_2}, \label{D1} \\
D_2 & = \Phi_{2,q}^2 (\kperp) \Phi_{1,p}(\pperp - \kperp) D_{A_1}, \label{D2} \\
D_{A_{2(1)}} & =  \Phi_{2(1),q}(\qperp + \kperp) + \Phi_{2(1),q}(\qperp - \kperp). \label{DA12}
\end{align}
The double-gluon correlation function $C_2(\bp,\bq)$ in Eq.~(\ref{C_2}) has recently been used to explain the ridges seen in the di-hadron correlations in p--p ($\sqrt{s}=7\tev$) and p--Pb ($\sqrt{s_{NN}}=5.02\tev$) collisions at LHC 
\cite{Dumitru:2010iy,Dusling:2012iga,Dusling:2012cg,Dusling:2012wy,Dusling:2013oia,Venugopalan:2013cga}.

In order to obtain ridges, one needs to supply $C_2(\bp,\bq)$ with UGDs that are approximately bell-shaped curves which peak around the saturation scale $p_\perp \sim \Qs$. So far, we have not yet mentioned which UGDs could be used for $C_2(\bp,\bq)$. The derivation of $C_2(\bp,\bq)$ is indeed independent of which UGD one wants to use as long as they have the aforementioned features.
We will utilize the UGDs evolved by the rcBK equation; they include 
gluon saturation and are roughly bell-shaped curves with their peaks around $\Qs$ (see Ref. \cite{Dusling:2012wy}). 

We find that $C_2(\bp,\bq)$ peaks at the azimuthal angle differences $\Delta \phi \sim 0$ and $\Delta \phi \sim \pi$
\be
C_2(\Delta \phi) \propto c + \cos(2 \Delta \phi),
\label{cos2phi}
\ee
where $c$ is a constant. How $\cos(2 \Delta \phi)$ structure (``double ridge'') arises from the glasma  can be easily seen as follows. For this proof-of-concept calculation, instead of realistic rcBK UGDs, we shall use a mock 
Gaussian-shaped UGD
\begin{figure}[t]
\includegraphics[scale=0.65]{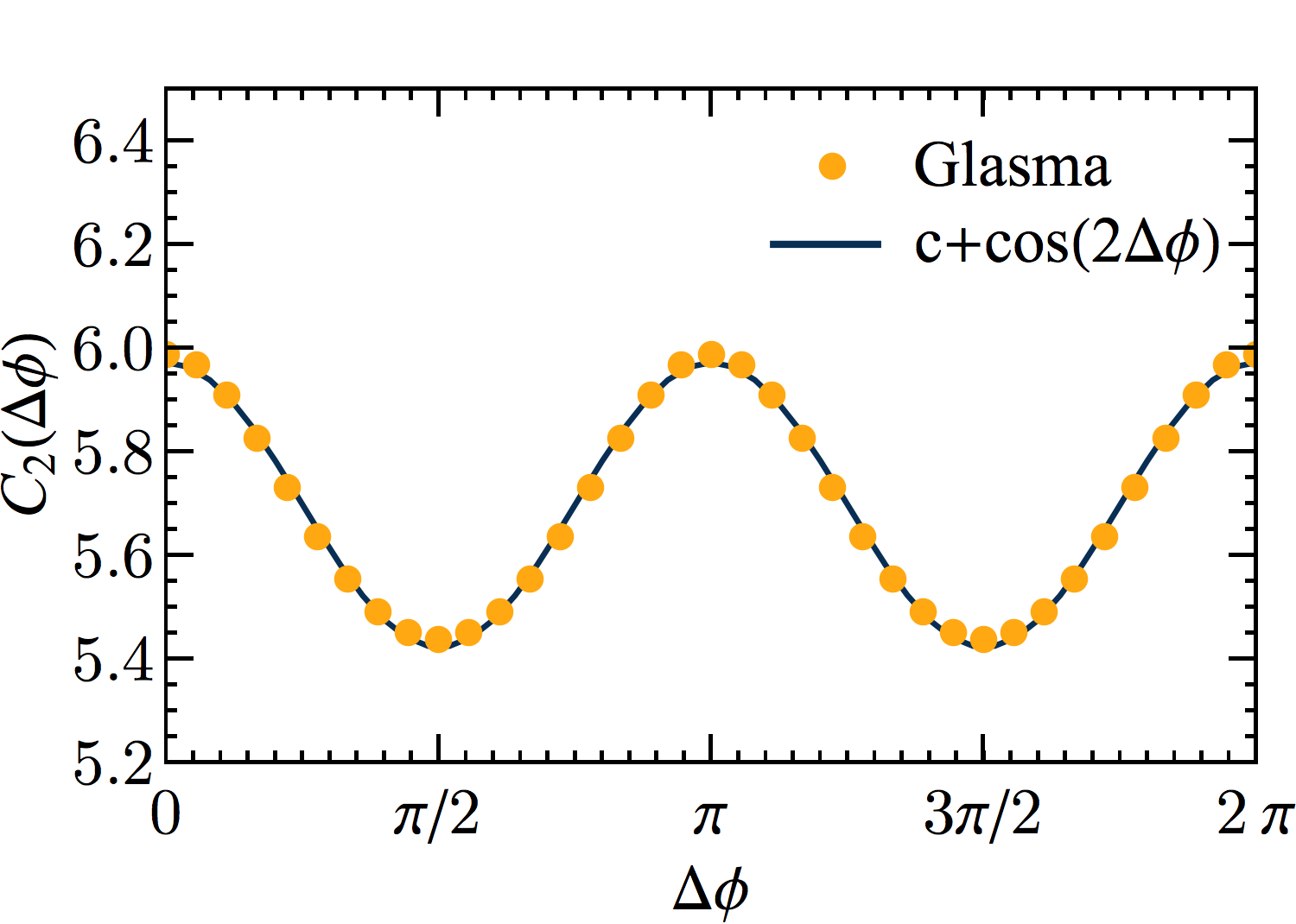}
\caption{
(Color online) Our numerical calculations of $C_2(\Delta \phi)$ from glasma diagrams supplied with the mock UGD given in Eq.~(\ref{mockUGD}) fit well to the function $c+\cos(2\Delta \phi)$ where $\Delta \phi$ is the azimuthal angle difference between the two produced gluons. How this structure emerges in the data can be seen in Fig.~\ref{fig:double-ridge}. The symmetric shape of the correlation function that peaks at $\Delta \phi \sim 0$ and $\Delta \phi \sim \pi$ is due to the interference of the Gaussian-shaped UGDs of the target and projectile. 
}
\label{fig:dipole}
\end{figure}
\be
\Phi(\pperp) \equiv \exp \left[ {-(p_\perp  - \Qs)^2} \right],
\label{mockUGD}
\ee
and we ignore the rapidity variable for the moment.
We also rewrite the expressions in Eqs.~(\ref{D1}),~(\ref{D2}) and (\ref{DA12}) in terms of azimuthal angles by using
\begin{align}
| \pperp - \kperp | & = \sqrt{p_\perp^2 + k_\perp^2 - 2 p_\perp k_\perp \cos(\phi_p-\phi_k)}, \\
| \qperp \pm \kperp | & = \sqrt{q_\perp^2 + k_\perp^2 \pm 2 q_\perp k_\perp \cos(\phi_q-\phi_k)}, \label{phiq}
\end{align}
where $p_\perp=|\pperp|$. The integration variable in Eq.~(\ref{C_2}) can be written as $d\phi_k d k_\perp k_\perp$. Since our mock UGD given in Eq.~(\ref{mockUGD}) peaks around $\Qs$, we can make the substitution ${p_\perp \sim q_\perp \sim k_\perp \sim \Qs}$ and get rid of the $d k_\perp$ integration. We ignore the rapidity coordinates $y_p$ and $y_q$ for the moment, and assume $A_1=A_2$. The azimuthal angles $\phi_p$ and $\phi_q$ are of the produced gluons whereas $\phi_k$ is to be integrated over. We define $\Delta \phi \equiv \phi_q - \phi_p$. We can rearrange our azimuthal coordinate system so that $\phi_p=0$, therefore the replacement $\phi_q=\Delta \phi$ can be made in Eq.~(\ref{phiq}). The only degree of freedom left is  $\Delta \phi$.  After carrying out the $\phi_k$ integration numerically for a list of values $\Delta \phi \in [0,2\pi]$, we verify the result given in Eq.~(\ref{cos2phi}) (see Fig.~\ref{fig:dipole}). This $\cos(2\Delta \phi)$ structure has been obtained numerically with rcBK UGDs in Refs. \cite{Dusling:2012cg,Dusling:2012wy,Dusling:2013oia}. However, we note in passing that $\cos(2 \Delta \phi)$ is not the only harmonic that exists in $C_2$. Although it is the dominant mode if the momenta of the tagged gluons are around $\Qs$, different structures become visible as the transverse momenta and rapidities of the gluons are varied.

\begin{figure}[t]
$\vcenter{\hbox{\includegraphics[scale=0.40]{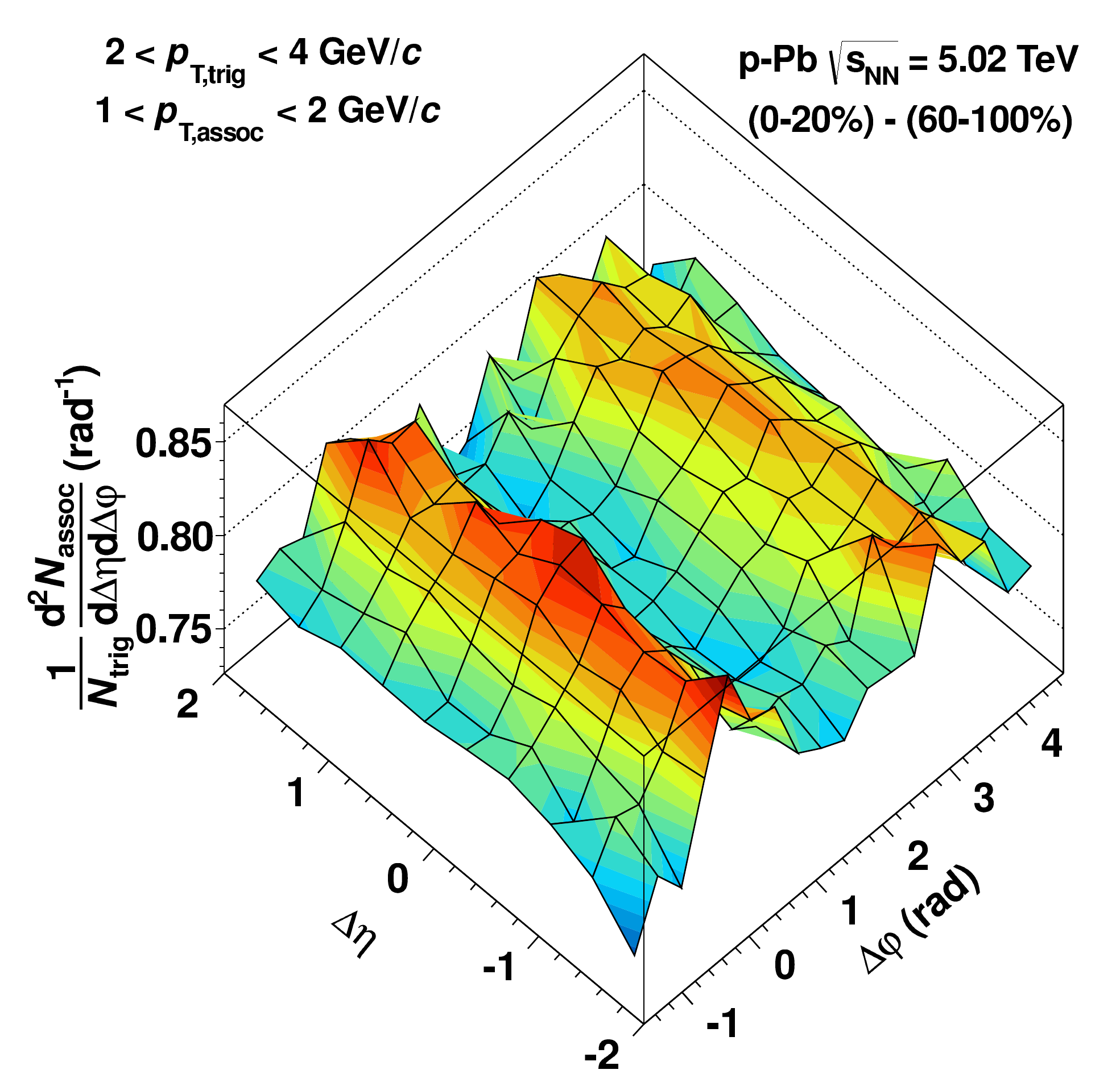} }}$
\hspace{1cm}
$\vcenter{\hbox{\includegraphics[scale=0.40]{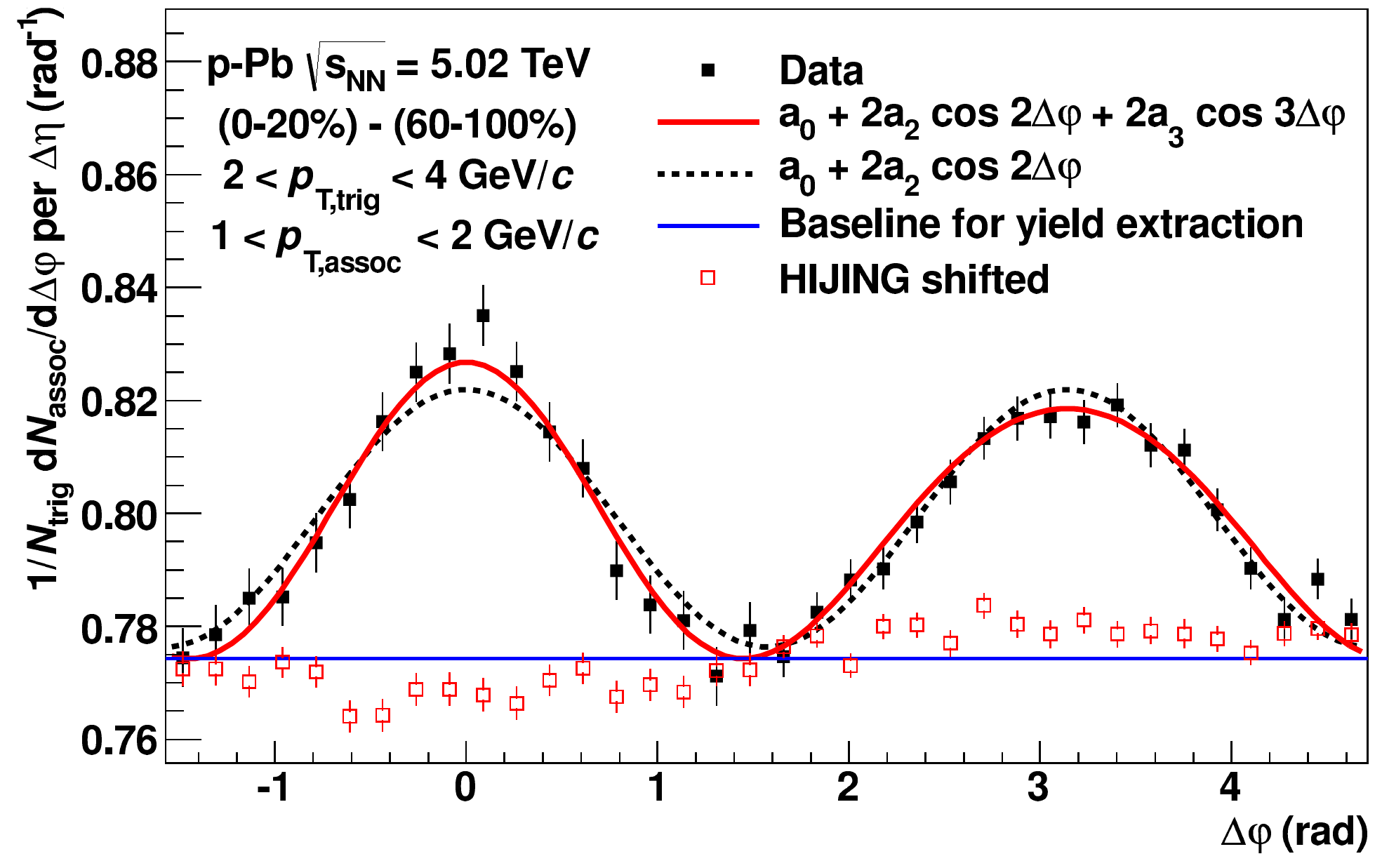} }}$
\caption{
(Color online) Symmetric, double ridge after the jet and resonance contributions are subtracted in the data \cite{Abelev:2012ola}.
}
\label{fig:double-ridge}
\end{figure}

The away side ridge $(\Delta \phi \sim \pi)$ has contributions also from jet fragmentation and resonance decays. In Ref.~\cite{Abelev:2012ola}, these contributions are removed and the double ridge structure became apparent in the data on p--Pb collisions  $\sqrt{s}=5.02\tev$. It has been observed that the near-side ridge ($\Delta \phi \sim 0$) was always accompanied with an identical ridge on the away side ($\Delta \phi \sim \pi$) (see Fig.~\ref{fig:double-ridge}), and this observation did not change with the systematics of event class and $p_T$ intervals of the measured hadrons. Also, based on the fact that the subtraction process is mostly independent of $p_T$, it has been concluded that the away-side ridge did not solely originate from the jet physics even though it included 
the jet and resonance decay contributions.

Although the glasma model is remarkably successful in explaining the ridges and reproducing their systematics in $N_{\rm track}$ and $p_T$ windows, there are other competing explanations which utilize the idea of final state collectivity (such as hydrodynamics) for p--p and p--Pb collisions, which has hitherto been thought to be impractical. At the moment, the di-hadron correlations are not enough to distinguish uniquely between the two scenarios leading to the ridges; final state collectivity or initial state effects by glasma. In order to settle this outstanding question, we suggest that one should look for possible ridges in triple-hadron, quadruple-hadron or even higher order hadron correlations. In the next section, we shall derive the triple-gluon $C_3(\bp,\bq,\bl)$ and quadruple-gluon $C_4(\bp,\bq,\bl,\bw)$ correlation functions for this purpose.

%%%%%
%%%%%
%%%%%
\section{Triple-Gluon Inclusive Distribution from Glasma}

In this section, we calculate the triple-gluon correlation function at arbitrary rapidity and transverse momentum\footnote{The triple-gluon correlation function that has been derived in Ref. \cite{Dusling:2009ar} did not have rapidity and transverse momentum dependences since the charge densities $\mu^2_{A_{1,2}}$ therein have been taken to be constant. Hence, it is not suitable for 
calculating ridges in p--p and p--Pb collisions without resorting to collective flow in hydrodynamics sense.}. Azimuthal collimations of ridges in tri-hadron and quadro-hadron correlations can be calculated from them.

\begin{figure}[t]
$\vcenter{\hbox{\includegraphics[scale=0.65]{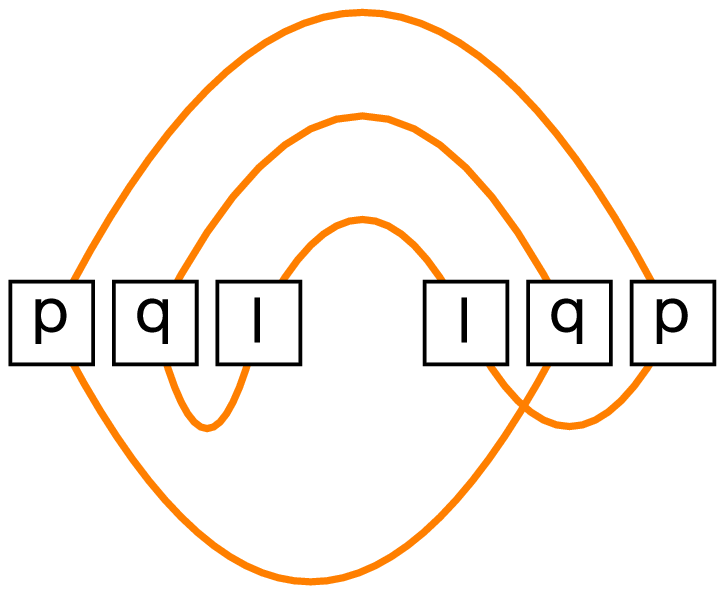}}}$
\hspace{1cm}
$\vcenter{\hbox{ \includegraphics[scale=0.9]{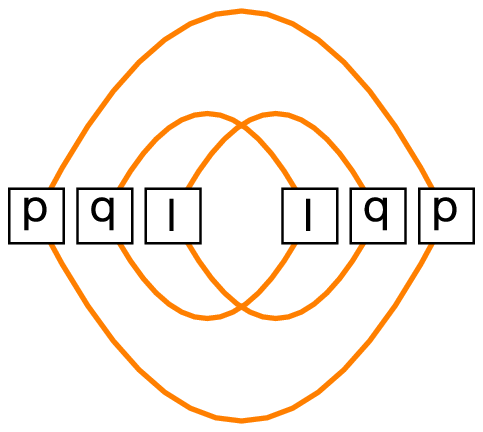}}}$
\caption{Examples of (left) connected and (right) disconnected triple-gluon glasma diagrams contributing to ${\cal F}^{(3)}$ in Eq.~(\ref{calF}). The disconnected diagram can be written as multiplication of single- and double-gluon distributions; it is not calculated since it is already subtracted from $C_3(\bp,\bq,\bl)$.}
\label{fig:triple}
\end{figure}

The triple-gluon inclusive distribution can be written as
\be
\left<  \frac{d N_3}{d^2 \bo{p}_\perp d y_p d^2 \bq_\perp d y_q d^2 \bl_\perp d y_l} \right>_\LLog  = \frac{1}{2^3 (2 \pi)^9} \sum_{\lambda \lambda^\prime \lambda^{\prime \prime}} \left<  | {\cal M}^{a a^\prime a^{\prime \prime}}_{\lambda \lambda^\prime \lambda^{\prime \prime}} (\bp,\bq,\bl) |^2  \right>,
\label{dN3}
\ee 
where 
\be
{\cal M}^{a a^\prime a^{\prime \prime}}_{\lambda \lambda^\prime \lambda^{\prime \prime}} = p^2 q^2 l^2 A^{a,\mu}(p) A^{a^\prime,\nu}(q) A^{ a^{\prime \prime},\sigma}(l)  \epsilon_\mu^{(\lambda)}(\bo{p}) \epsilon_\nu^{(\lambda^\prime)}(\bo{q}) \epsilon_\sigma^{(\lambda^{\prime \prime})}(\bo{l}). 
\ee
By using Eq.~(\ref{Aexpn}), we can write Eq.~(\ref{dN3}) in a compact way
\begin{align}
\left<  \frac{d N_3}{d^2 \bo{p}_\perp d y_p d^2 \bq_\perp d y_q d^2 \bl_\perp d y_l} \right>_\LLog  & =    \frac{-1}{2^3 (2 \pi)^9}  (g^3)^6 f^{abc} f^{afg} f^{ a^\prime de} f^{ a^\prime hi} f^{ a^{\prime \prime}mr} f^{ a^{\prime \prime}ns} \nonumber \\
& \quad \times \int \left( \prod^3_{i=1}  \frac{d \bo{k}_{2i-1 \perp}}{(2\pi)^2} \frac{d \bo{k}_{2i \perp}}{(2\pi)^2} 
\frac{L^\mu(\bo{r}_{i \perp},\bo{k}_{2i-1 \perp}) L_\mu(\bo{r}_{i \perp},\bo{k}_{2i \perp})}{ \bo{k}^2_{2i-1 \perp} (\bo{r}_{i \perp}- \bo{k}_{2i-1 \perp})^2 \,\, \bo{k}^2_{2i \perp}  (\bo{r}_{i \perp}- \bo{k}_{2i \perp})^2}  \right) {\cal F}^{(3)},
\end{align}
where $\bo{r}_{i \perp}=(\pperp,\qperp,\lperp)$ and
\begin{align}
{\cal F}^{(3)} \equiv  \Big< &
\rhot_1^b (\bo{k}_{1 \perp},y_p) \rhot_2^c (\pperp - \bo{k}_{1 \perp},y_p) 
\rhot_1^{*f} (\bo{k}_{2 \perp},y_p) \rhot_2^{*g} (\pperp - \bo{k}_{2 \perp},y_p)\nonumber \\
&
\, \times \rhot_1^d (\bo{k}_{3 \perp},y_q) \rhot_2^e (\qperp - \bo{k}_{3 \perp},y_q) 
\rhot_1^{*h} (\bo{k}_{4 \perp},y_q) \rhot_2^{*i} (\qperp - \bo{k}_{4 \perp},y_q)\nonumber \\
&
\, \times \rhot_1^m (\bo{k}_{5 \perp},y_l) \rhot_2^r (\lperp - \bo{k}_{5 \perp},y_l) 
\rhot_1^{*n} (\bo{k}_{6 \perp},y_l) \rhot_2^{*s} (\lperp - \bo{k}_{6 \perp},y_l)
\Big>.
\label{calF}
\end{align}
Here $\rhot_1$'s and $\rhot_2$'s can be contracted only among each other; i.e., there are no correlations between the charge densities of nucleus 1 (projectile) and nucleus 2 (target). For each of the projectile and target, there are $(2\times3-1)!!=15$ contractions; hence, 225 diagrams in total.
Examples of connected and disconnected diagrams are presented in Fig.~\ref{fig:triple}.
Here we shall only calculate the connected diagrams that are not power suppressed due to the contractions leading to extra delta functions, and at the same time non-vanishing in the limit of $\pperp,\qperp,\lperp\ll \Qs$ (see Refs. \cite{Dusling:2009ni,Dusling:2009ar}). These are the rainbow diagrams where dimers of the same momentum are contracted among each other either on nucleus 1 or nucleus 2 [see Fig.~\ref{fig:triple}(left)].
The disconnected diagrams are subtracted in the triple-gluon correlation 
function
\begin{align}
C_3(\bp,\bq,\bl)= & \left< \frac{d^3 N_3}{d^2 \bo{p}_\perp d y_p d^2 \bq_\perp d y_q d^2 \bl_\perp d y_l} \right>
-\left<  \frac{d^2 N_2}{d^2 \bo{p}_\perp d y_p d^2 \bq_\perp d y_q} \right>_c \, \left< \frac{dN_1}{d^2 \bl_\perp d y_l} \right> \nonumber \\
& -\left<  \frac{d^2 N_2}{d^2 \bo{p}_\perp d y_p d^2 \bl_\perp d y_l} \right>_c \, \left< \frac{dN_1}{d^2 \bq_\perp d y_q} \right>
-\left<  \frac{d^2 N_2}{d^2 \bo{l}_\perp d y_l d^2 \bq_\perp d y_q} \right>_c \, \left< \frac{dN_1}{d^2 \bp_\perp d y_p} \right>  \nonumber \\
& - \left< \frac{dN_1}{d^2 \bp_\perp d y_p} \right> \left< \frac{dN_1}{d^2 \bq_\perp d y_q} \right> \left< \frac{dN_1}{d^2 \bl_\perp d y_l}  \right>. \label{connected}
\end{align}
Here $C_3(\bp,\bq,\bl)$ is the third cumulant, the first term on the right-hand side is the third moment, and the terms with the subscript ``c'' are the second cumulants, which are $C_2$'s. The cumulants include only the connected diagrams whereas the moments include both the connected and disconnected diagrams\footnote{Alternatively, one can use the second moment rather than the second cumulant in the expansion of $C_3(\bp,\bq,\bl)$ as done in Ref.~\cite{Dusling:2009ar}, and write ${C_3(\bp,\bq,\bl)= \left< d^3 N_3\right>
-3 \left< d^2 N_2 \right> \left< dN_1\right> 
+2 \left< dN_1\right> \left< dN_1\right> \left< dN_1\right>} $. The reason why we prefer the expansion in terms of cumulants rather than moments will be clear when we calculate the quadruple-gluon inclusive distribution. At that order, the naive cumulant expansion should be slightly modified so that the subtracted disconnected diagrams do not include the irrelevant diagrams which mix an upper and a lower disconnected two-gluon rainbow diagrams. These irrelevant diagrams are only partially in a rainbow form.}. 
Let us first find out how many connected and disconnected rainbow diagrams the third moment has. There are $(2\times3-1)!!=15$ contractions for each of the upper rainbow and lower rainbow. Both of these include the maximally disconnected diagram that is formed by three concentric circles. By avoiding double counting, we find there are 29 diagrams (connected and disconnected) in the third moment.
We also know from the previous section that the second cumulant includes 4 diagrams. From Eq.~(\ref{connected}), we can find the total number of diagrams included in $C_3(\bp,\bq,\bl)$
\be
29-3\times4-1=16.
\ee
Since the experimentally relevant quantity is $C_3(\bp,\bq,\bl)$, we will only calculate these 16 diagrams that will lead us to $C_3(\bp,\bq,\bl)$.

In order to write ${\cal F}^{(3)}$ in terms of $\mu^2_{A_1,A_2}$, we shall use the correlation functions given in Eqs.~(\ref{corr1again}), (\ref{corr2}) and (\ref{corr3}) by keeping in mind that we now have one more rapidity variable $y_l$. The rapidity structure of these correlation functions are as follows. For correlations in $A_1$, $y_p$ will be chosen over $y_q$ and $y_l$, and $y_q$ will be chosen over $y_l$. For contractions in $A_2$, this ordering is reversed. Hence, for example
\begin{align}
\left< \rhot_1(y_p) \rhot_1(y_q) \right>  \propto \mu^2_{A_1}(y_p), \label{rapidity-pattern1} \\
\left< \rhot_1(y_q) \rhot_1(y_l) \right>  \propto \mu^2_{A_1}(y_q),\\
\left< \rhot_2(y_p) \rhot_2(y_l) \right>  \propto \mu^2_{A_2}(y_l). \label{rapidity-pattern}
\end{align}
This pattern arises from our convention that the gluon with momentum $\bo{p}$ is close to the projectile, the one with momentum $\bo{l}$ is close to the target, and the one with momentum $\bo{q}$ is in between. Therefore, the rapidity ordering is $y_p>y_q>y_l$ (see also \cite{Dusling:2009ni}).

After this point, the calculation proceeds similar to the cases of the single- and double-gluon calculations outlined before. The contractions between the color structure functions and the delta functions involving color factors give rise to a term $N^3_c(N^2_c-1)$.
We skip the details of the tedious but straightforward calculations and show the final form of the triple-gluon correlation function\footnote{Equation~(12) in Ref.~\cite{Gelis:2009wh} does not reproduce
correctly neither the prefactor of Eq.~(\ref{C_3}) here nor the momentum dependence of the UGDs in Eq.~(\ref{T12}).
}
\be
C_3(\bp,\bq,\bl) = \frac{ \as^3  N_c^3 S_\perp}{ \pi^{12}  (N_c^2-1)^5}
\frac{1}{\pperp^2 \qperp^2\lperp^2} \int \frac{d^2 \kperp}{(2\pi)^2}  ({\cal T}_1+{\cal T}_2),
\label{C_3}
\ee
where
\begin{align}
{\cal T}_1 = & 2 \times \left( \Phi_{1,p}(\kperp) \right)^2  \Phi_{1,q}(\kperp) \Phi_{2,p}(\pperp - \kperp) {\cal T}_{A_2 },\label{calT1} \\
{\cal T}_2 = &  2 \times \left( \Phi_{2,l}( \kperp ) \right)^2  \Phi_{2,q}( \kperp ) \Phi_{1,p}( \pperp - \kperp ) {\cal T}_{A_1 }, \label{calT2} \\
{\cal T}_{A_1,A_2} = &  \left[ \Phi_{1(2),q}( \qperp-\kperp) +   \Phi_{1(2),q}( \qperp+\kperp)    \right]
 \left[ \Phi_{1(2),l}( \lperp-\kperp) +   \Phi_{1(2),l}( \lperp+\kperp)    \right].  \label{T12}
\end{align}
The prefactor in Eq.~(\ref{C_3}) follows a pattern in accordance with Eqs.~(\ref{dN1result}) and (\ref{C_2}). Note that we added the factor of two into the definitions in Eqs.~(\ref{calT1}) and (\ref{calT2}). This factor of two arises from the diagrams which are mirror images of each other in the left-right direction. These pair of distinct diagrams, however, contribute precisely the same way. 

%%%%%
%%%%%
%%%%%
\section{Quadruple-Gluon Inclusive Distribution from Glasma}

\begin{figure}[t]
\begin{centering}
$\vcenter{\hbox{\includegraphics[scale=0.80]{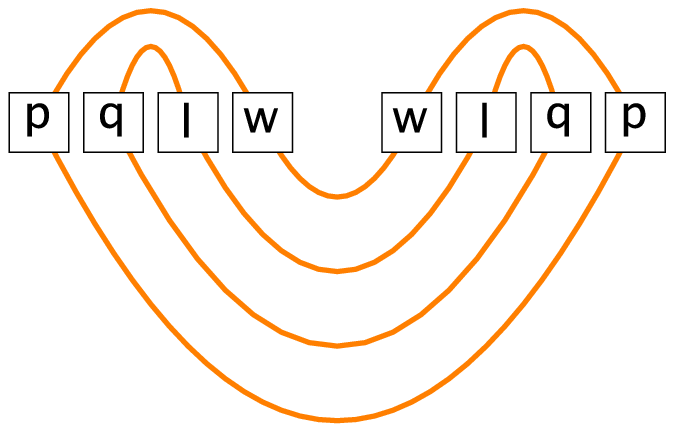}}}$
\hspace{1cm}
$\vcenter{\hbox{\includegraphics[scale=0.80]{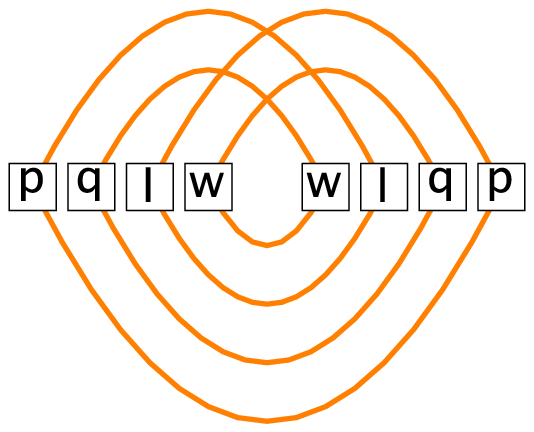}}}$
\par\end{centering}
\caption{Disconnected glasma diagrams for quadruple-gluon correlations, which are already subtracted in $C_4(\bp,\bq,\bl,\bw)$.}
\label{fig:quadro}
\end{figure}

In this section, we calculate the quadruple-gluon correlation function at arbitrary rapidity and transverse momentum.
The procedure is identical to that of the triple-gluon calculation. At this order, there are $(2\times4-1)!!\times(2\times4-1)!!=11025$ diagrams in total. Some of the disconnected diagrams are shown in Fig.~\ref{fig:quadro}. We will only calculate the rainbow diagrams as mentioned in the previous section, and there are 96 of them at this order. This can be verified as follows. We first expand the fourth cumulant in terms of the third, second and first cumulants
\begin{align}
C_4(\bp,\bq,\bl,\bw)= & \left< d^4 N_4\right> -4 \left< d^3 N_3\right>_c \left< d N_1\right> - 3\left< d^2 N_2\right>_c \left< d^2 N_2\right>_c \nonumber \\ 
& \quad -6 \left< d^2 N_2\right>_c \left< d N_1\right> \left< d N_1\right> - \left< d N_1\right> \left< d N_1\right> \left< d N_1\right> \left< d N_1\right>, \label{4thCumulant}
\end{align}
where we used a compact notation; compare Eq.~(\ref{4thCumulant}) with Eq.~(\ref{connected}). The forth moment $\left< d^4 N_4\right>$ contains 
\be
2\times (2\times4-1)!! -1=209
\ee
connected and disconnected diagrams, where the subtracted ``1'' is for the maximally disconnected diagram
so that we do not double count it.
We know from the previous sections that the cumulant $\left< d^3 N_3\right>_c=C_3$ contains 16 diagrams, and the cumulant ${\left< d^2 N_2\right>_c=C_2}$ contains 4 diagrams. From Eq.~(\ref{4thCumulant}) we find the number of connected diagrams that $C_4$ contains 
\be
209-4\times16-3\times(2\times2+2\times2)-6\times4-1=96.
\ee
Although $\left< d^2 N_2\right>_c$ contains 4 diagrams, note that the term $\left< d^2 N_2\right>_c \left< d^2 N_2\right>_c$ in Eq.~(\ref{4thCumulant}) does not simply contribute $4\times4=16$. This would be a mistake since we would have counted the disconnected diagrams which would include an upper rainbow from one of $\left< d^2 N_2\right>_c$'s and a lower rainbow from the other one; these are not rainbow diagrams. This also tells us that the cumulant expansions of rainbow diagrams should be modified when necessary.

Since we already outlined the steps of the calculation of glasma diagrams in the previous sections, here we skip the details and only present the final result
\be
C_4(\bp,\bq,\bl,\bw) = \frac{\as^4  N_c^4 S_\perp}{ \pi^{16}  (N_c^2-1)^7}
\frac{1}{\pperp^2 \qperp^2\lperp^2 \wperp^2} \int \frac{d^2 \kperp}{(2\pi)^2}  ({\cal Q}_1+{\cal Q}_2),
\label{C_4}
\ee
where
\begin{align}
{\cal Q}_1 = & \left( \Phi_{1,p}(\kperp) \right)^2  \Phi_{1,q}(\kperp)  \Big[ 4 \times \Phi_{1,l}(\kperp) + 2 \times \Phi_{1,q}(\kperp) \Big]   \Phi_{2,p}(\pperp - \kperp) {\cal Q}_{A_2 },\label{Q1} \\
{\cal Q}_2 = & \left( \Phi_{2,w}( \kperp ) \right)^2   \Phi_{2,l}(\kperp) \Big[ 4 \times \Phi_{2,q}(\kperp) + 2  \times \Phi_{2,l}(\kperp) \Big]  \Phi_{1,p}( \pperp - \kperp ) {\cal Q}_{A_1 },\label{Q2} \\
{\cal Q}_{A_1,A_2} = &  \left[ \Phi_{1(2),q}( \qperp-\kperp) +   \Phi_{1(2),q}( \qperp+\kperp)    \right]
 \left[ \Phi_{1(2),l}( \lperp-\kperp) +   \Phi_{1(2),l}( \lperp+\kperp)  \right] \nonumber \\
 & \quad\quad\quad\quad\quad\quad \times  \left[ \Phi_{1(2),w}( \wperp-\kperp) +   \Phi_{1(2),w}( \wperp+\kperp)  \right] .
\end{align}
The factors of ``$4$'' and ``$2$'' in Eqs.~(\ref{Q1}) and (\ref{Q2}) arise from the complete or partial left-right mirror
symmetry among the glasma diagrams considered here.

%%%%%
%%%%%
%%%%%
\section{Plots for Triple- and Quadrupole-gluon Azimuthal Correlations}

\begin{figure}[t!]
\begin{centering}
$\vcenter{\hbox{\includegraphics[scale=0.7]{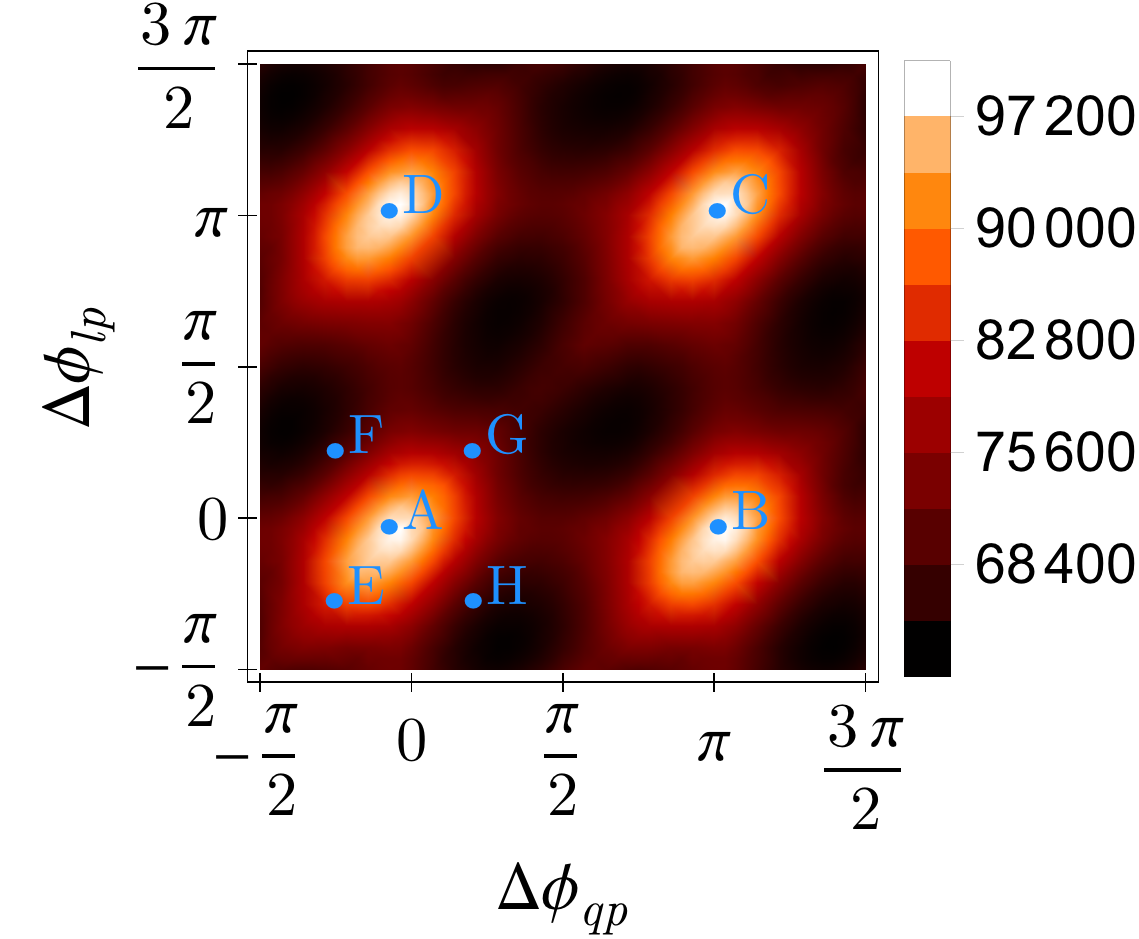}}}$
\hspace{1cm}
$\vcenter{\hbox{\includegraphics[scale=0.12]{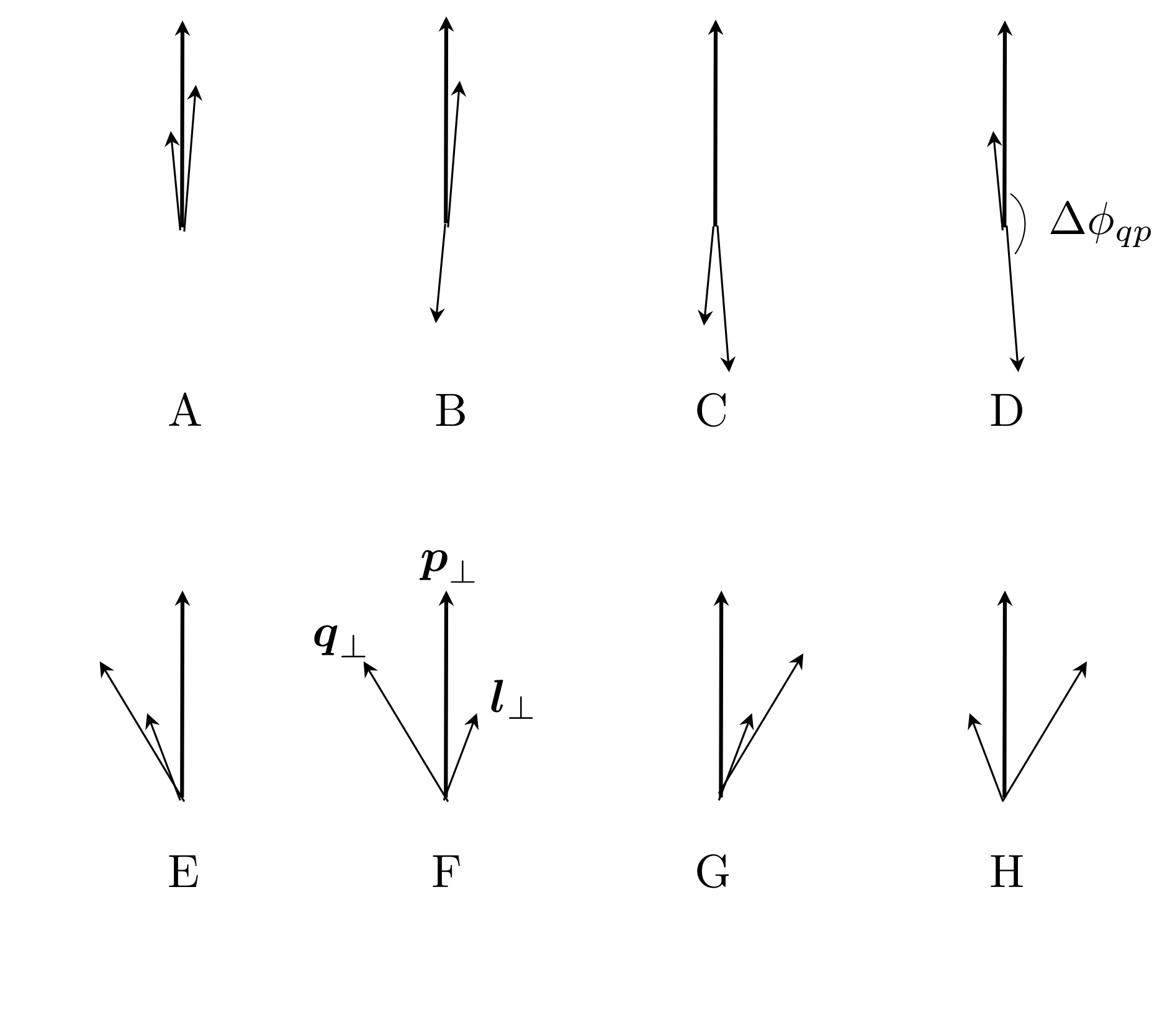}}}$
\par\end{centering}
\caption{(Color online) (left) Density plot of a typical triple-gluon azimuthal correlation function $C_3(\Delta \phi_{qp},\Delta \phi_{lp})$ in units of ${\as^3  N_c^3 S_\perp / \pi^{12}  (N_c^2-1)^5}$ [see Eq.~(\ref{C_3})]. (right) The relative azimuthal configurations of the three gluons for the points designated on the density plot on the left. The longest arrow represents $\pperp$, the middle-sized one is for $\qperp$ and the shortest arrow is for $\lperp$. We choose 
$\pperp$ as the trigger particle and arrange the transverse coordinate system such that its azimuthal position is fixed. Hence, density plots like the one on the left include all possible azimuthal configurations of $\qperp$ and $\lperp$ with respect to $\pperp$.
}
\label{fig:example}
\end{figure} 

\begin{figure}[b!]
\begin{centering}
\includegraphics[scale=0.332]{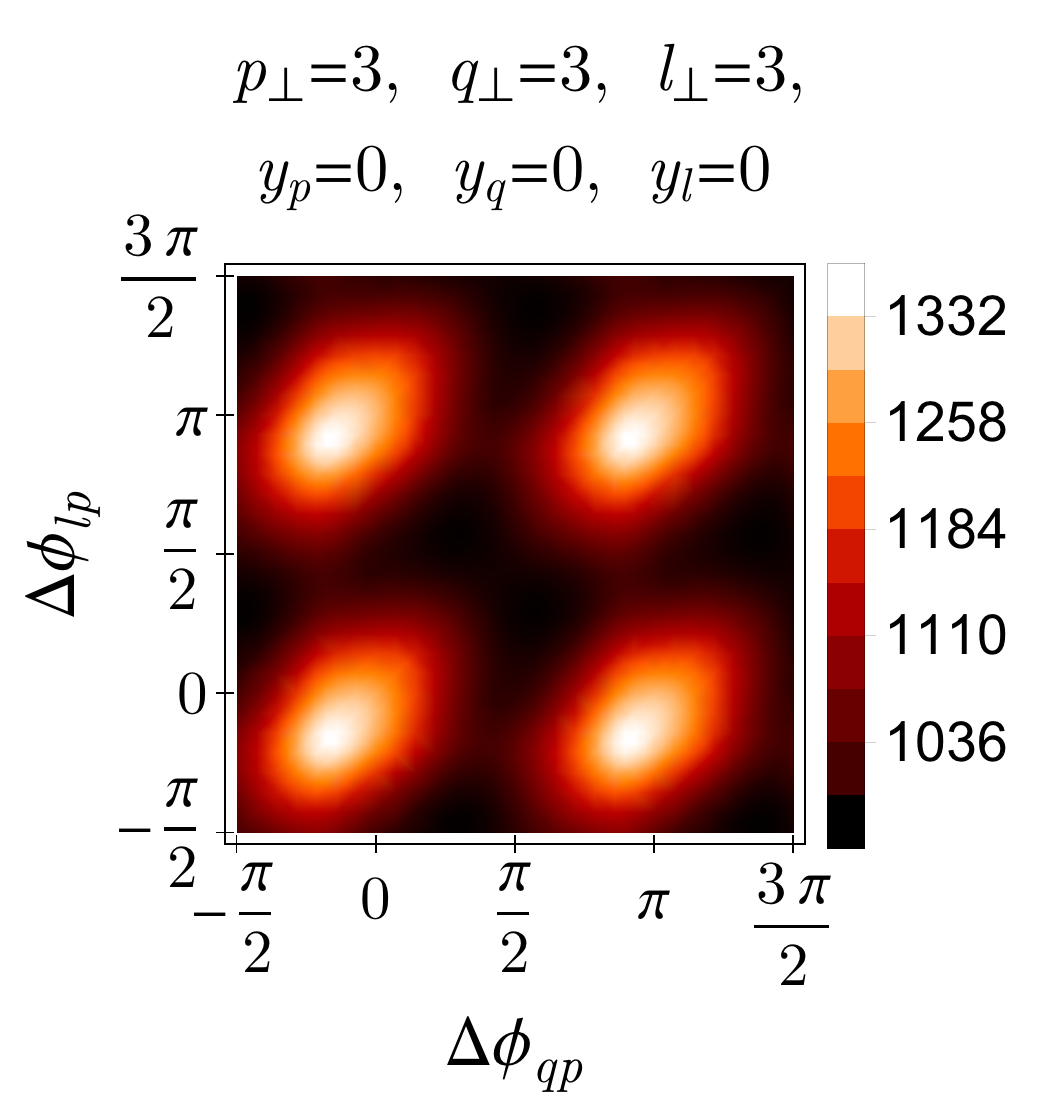}
\includegraphics[scale=0.322]{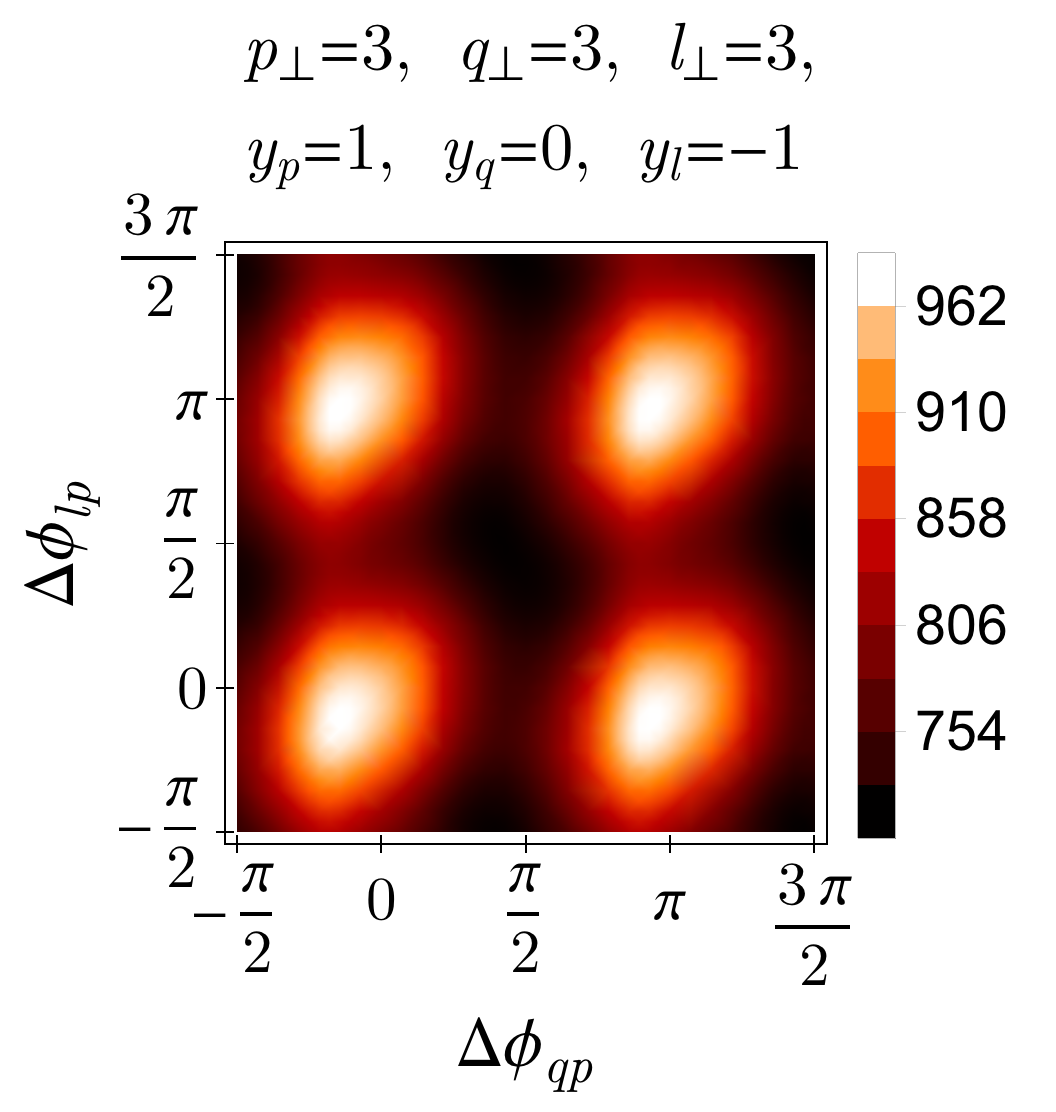}
\includegraphics[scale=0.322]{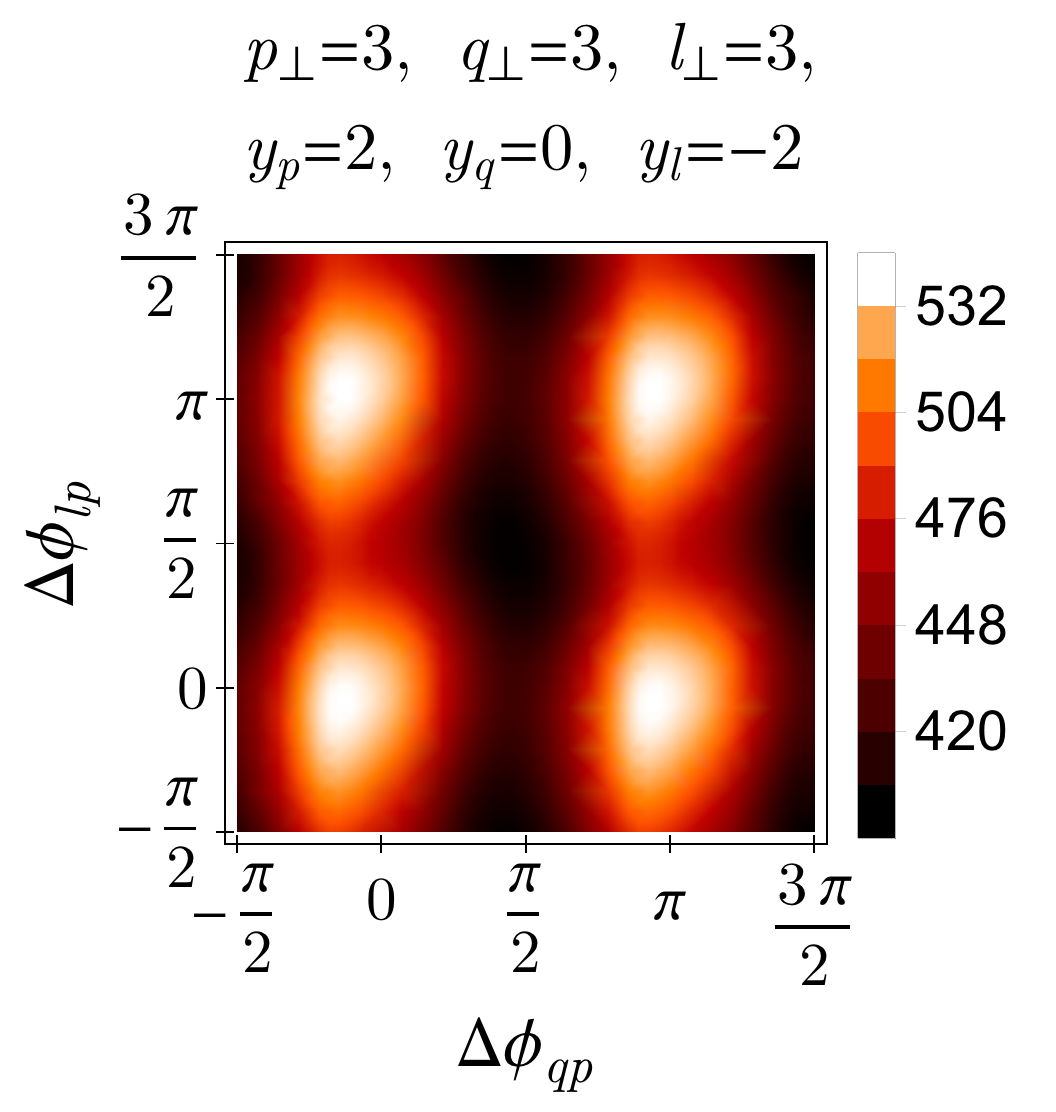}
\includegraphics[scale=0.332]{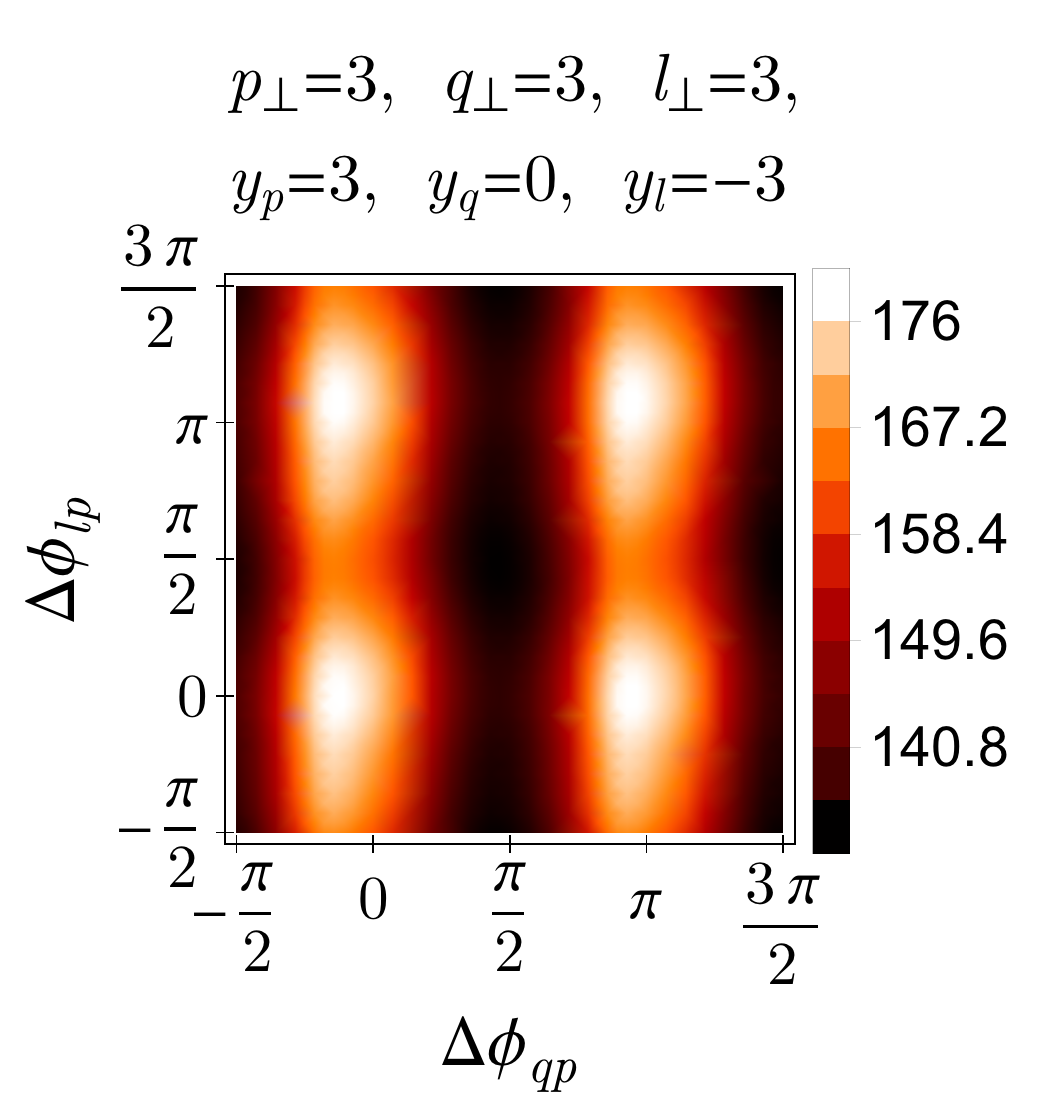}
\par\end{centering}
\caption{(Color online) Triple-gluon azimuthal correlation $C_3(\Delta \phi_{qp},\Delta \phi_{lp})$ for p--p collisions at $\sqrt{s}=7\tev$ in units of ${\as^3  N_c^3 S_\perp / \pi^{12}  (N_c^2-1)^5}$ [see Eq.~(\ref{C_3})]. The transverse momenta labels are in units of GeV. The magnitude of the correlation decreases with increasing rapidity gap between the gluons.}
\label{fig:ppSet1}
\end{figure}

%% Fig. 8
Figure~\ref{fig:ppSet1} shows triple-gluon azimuthal correlations with various rapidity configurations for p--p collisions at ${\sqrt{s}=7\tev}$. For both the projectile and target we used UGDs that were evolved with the initial scale $Q_0^2=0.168\gev^2$ (see Appendix \ref{appxUGD}). The correlation decreases as the rapidity separation between the gluons grows. Also, the shape of the signal is dependent on the separations of the gluons in rapidity.

%% Fig. 9
\begin{figure}[t!]
\begin{centering}
\includegraphics[scale=0.325]{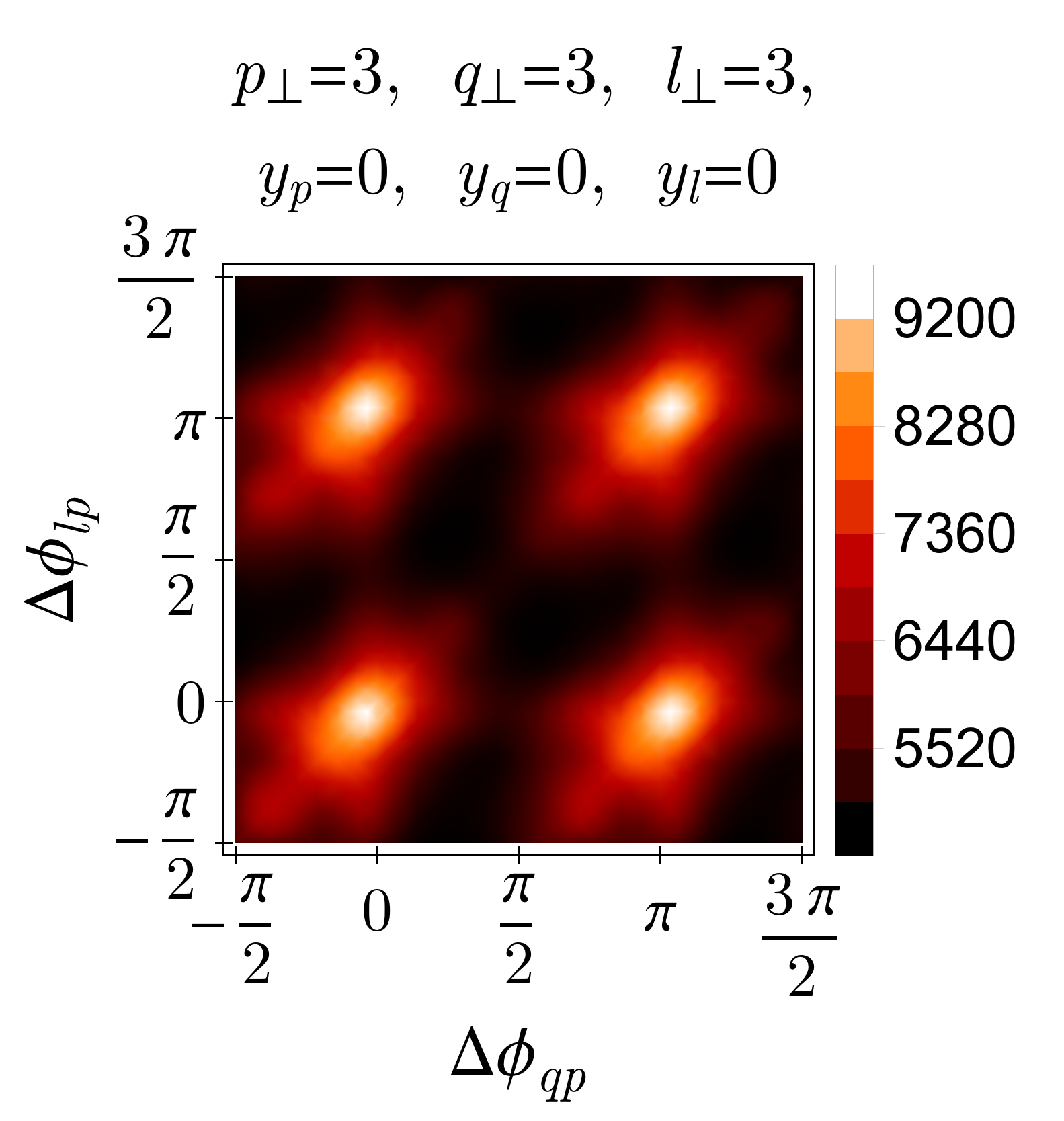}
\includegraphics[scale=0.325]{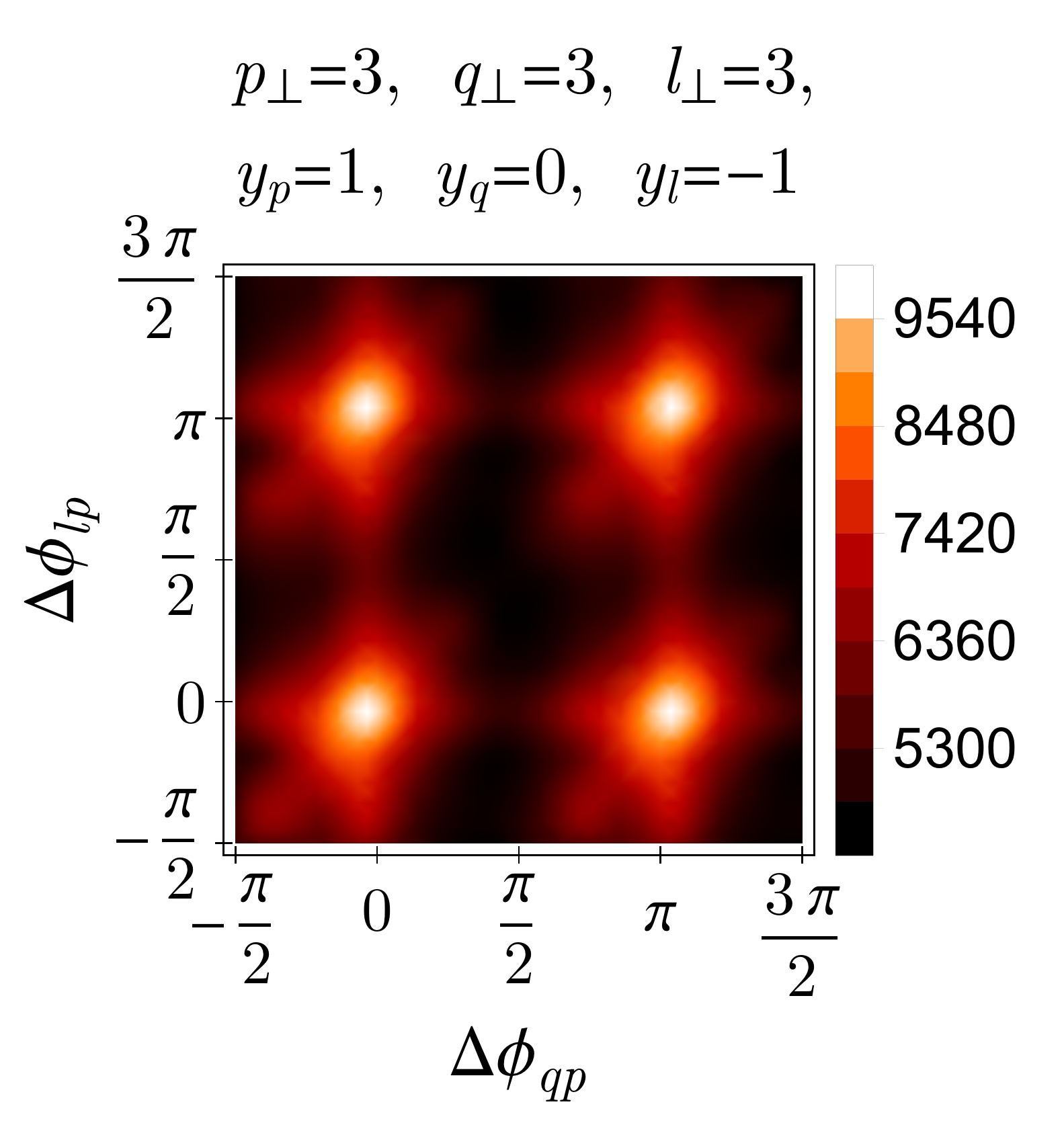}
\includegraphics[scale=0.325]{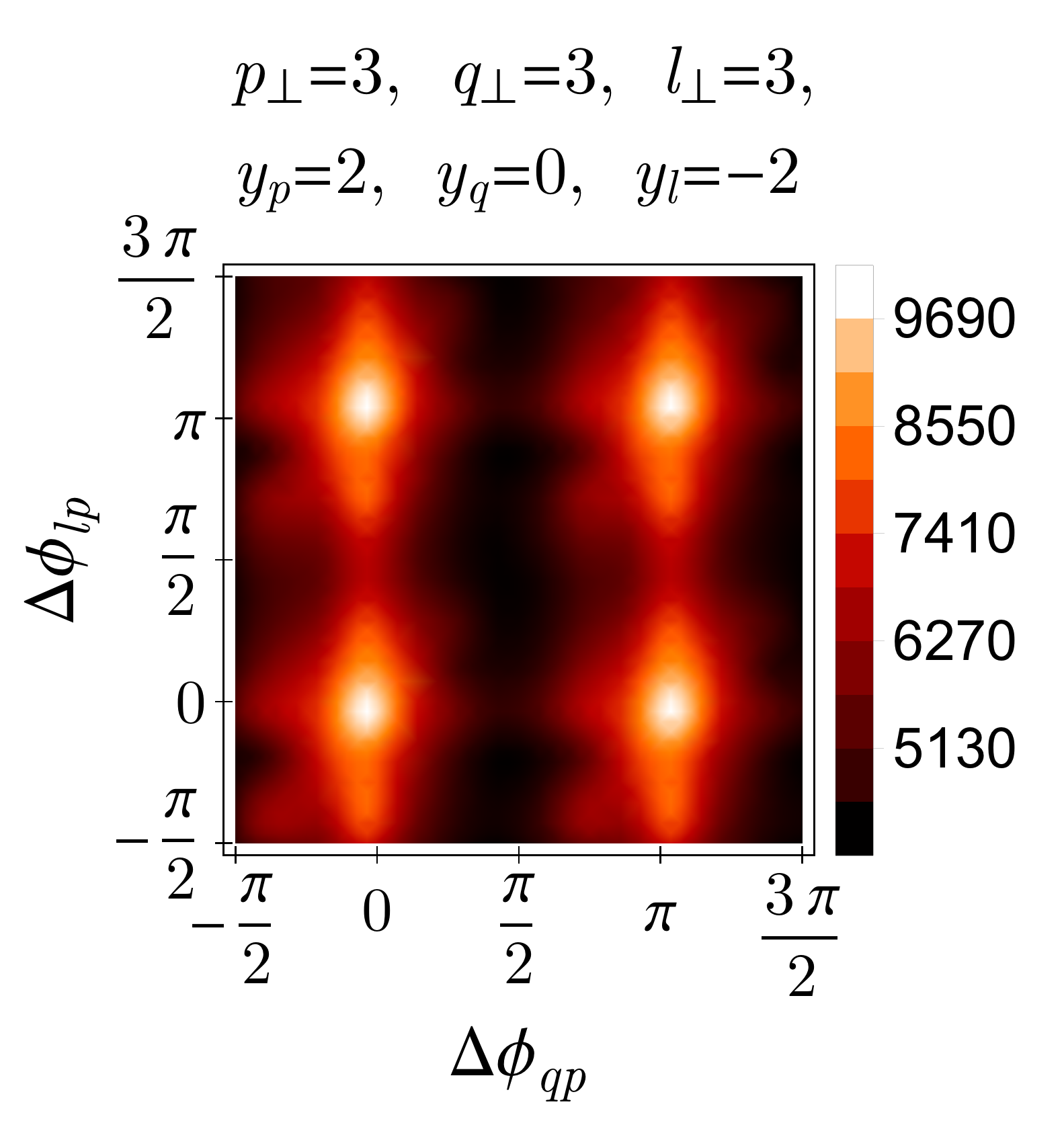}
\includegraphics[scale=0.325]{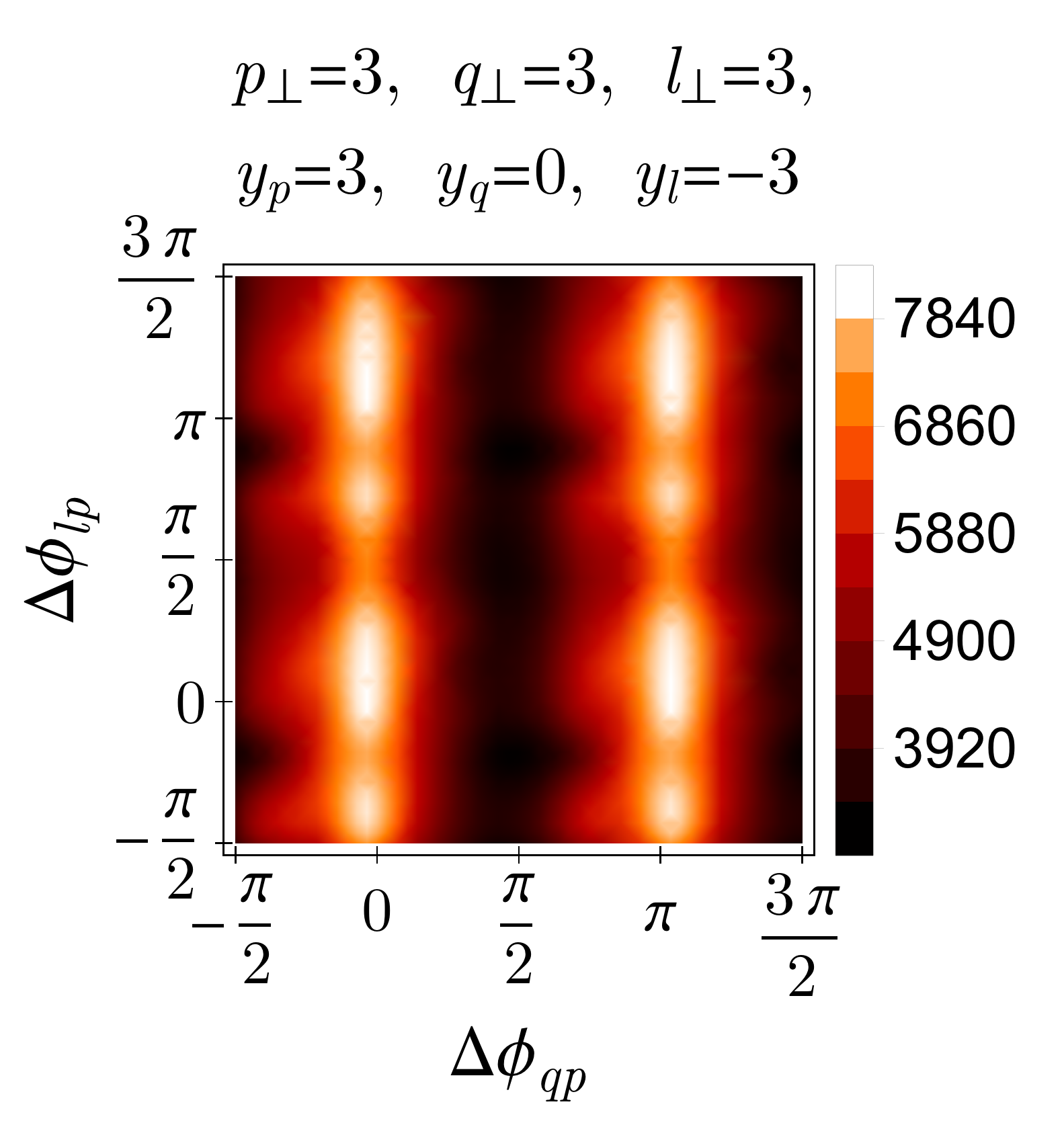}
\par\end{centering}
\caption{(Color online) Triple-gluon azimuthal correlation $C_3(\Delta \phi_{qp},\Delta \phi_{lp})$ for p--Pb collisions at $\sqrt{s}=5.02\tev$ in units of ${\as^3  N_c^3 S_\perp / \pi^{12}  (N_c^2-1)^5}$ [see Eq.~(\ref{C_3})]. The transverse momenta labels are in units of GeV. The magnitude of the correlation decreases with increasing rapidity gap between the gluons.}
\label{fig:pPbSet2}
\end{figure}

In this section we present some plots for triple- and quadrupole-gluon 
azimuthal correlations for different set of momenta, beam energy and 
number of participants in a collision. Also, we predict higher-dimensional
ridges in triple- and quadruple-hadron correlations  
for p--p and p--Pb collisions at LHC, which have yet to be measured.
In our density plots for triple-gluon correlations, it can be seen that the three gluons are 
collimated at the azimuthal angle $(0,0)$, and partially collimated at $(0,\pi)$, $(\pi,0)$ and 
$(\pi,\pi)$. This structure is preserved as the rapidity separation between the gluons increase.
As we have ``double ridges'' in double-gluon correlations, we find ``quadruple ridges'' in triple-gluon 
correlations, and ``octuple ridges'' in the quadruple-gluon correlations.

We use rcBK UGDs, and we
parametrize the large-$x$ contribution as done in \cite{Dusling:2009ni,Gelis:2006tb,Fujii:2006ab}.
The triple- and quadrupole-gluon azimuthal correlations should ultimately be convoluted with fragmentation functions
to be able to make quantitative predictions since the data is on the final
hadron distribution rather than gluons.
On the other hand, even though the ridge correlations are seen on the final hadron spectra,
the azimuthal correlations of gluons presented here already give
qualitative information about ridges and their systematics, i.e., how
the triple- and quadruple-hadron correlations would change 
depending on the momenta of hadrons and number of participants
in a collision. 

In double-gluon azimuthal correlations, there are one azimuthal angle difference and one rapidity difference. In the three-gluon case, there are two angle differences ($\Delta \phi_{qp}=\phi_q -\phi_p$ and $\Delta \phi_{lp}=\phi_l -\phi_p$ ) and two rapidity differences ($\Delta \eta_{pq}=\eta_p -\eta_q$ and  $\Delta \eta_{pl}=\eta_p -\eta_l$). 
On the quadruple-gluon level, there are three azimuthal angle differences and three rapidity differences. We take pseudorapidity ($\eta$) and rapidity ($y$) to be the same, and in our convention, the rapidity of the produced gluons are ordered as $y_p>y_q>y_l$ [see the discussion below Eq.~(\ref{connected})]. 

We plot $C_3(\Delta \phi_{qp},\Delta \phi_{lp},\Delta \eta_{pq},\Delta \eta_{pl})$ versus $\Delta \phi_{qp}$ and $\Delta \phi_{lp}$ for a given set of 
rapidity differences and gluons' momenta.
Various regions of a density plot of $C_3$ that correspond to different relative azimuthal configurations of the three gluons are explained in Fig.~\ref{fig:example}.

\begin{figure}[t]
\begin{centering}
\includegraphics[scale=0.363]{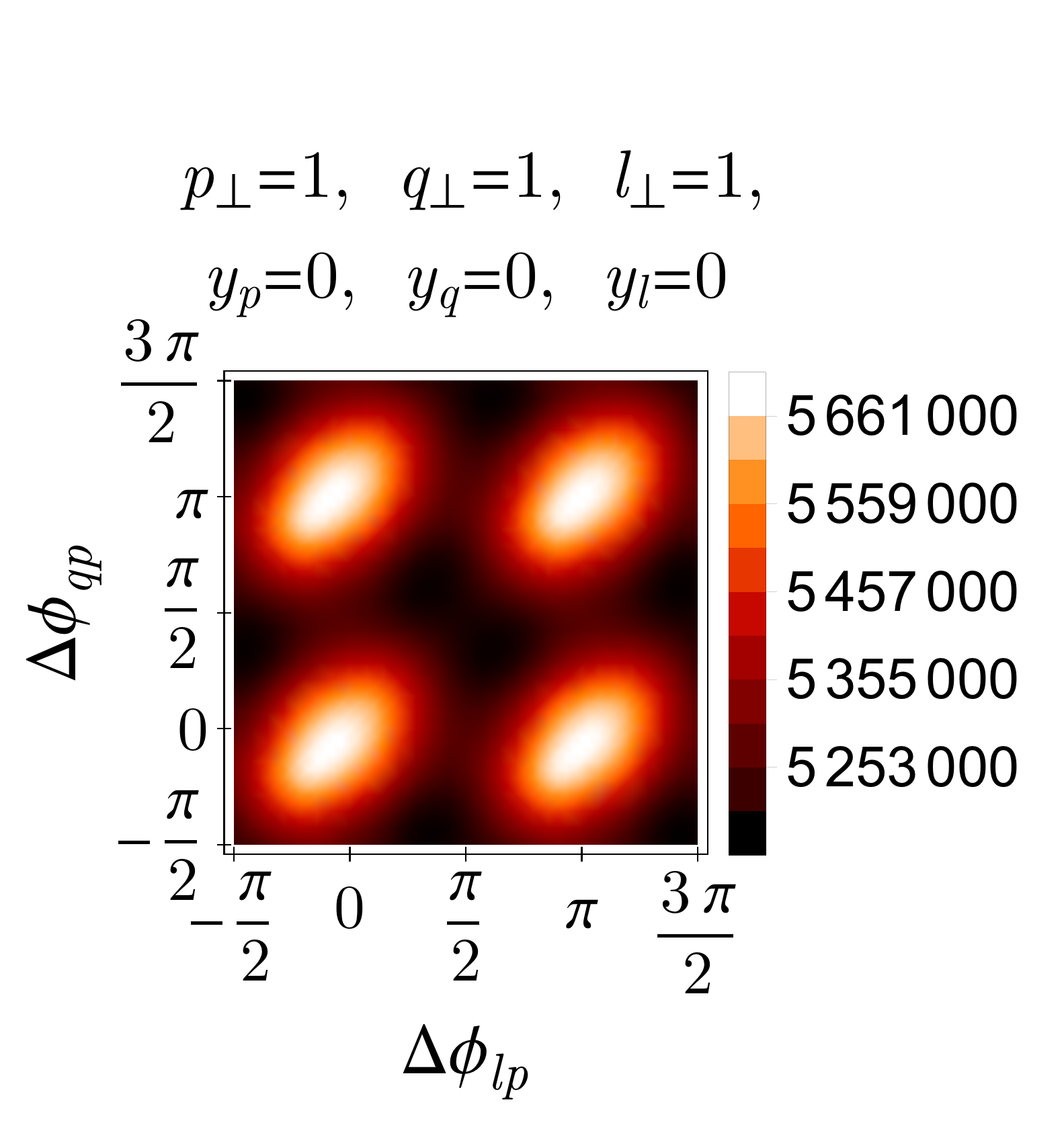} \hspace{-4mm}
\includegraphics[scale=0.335]{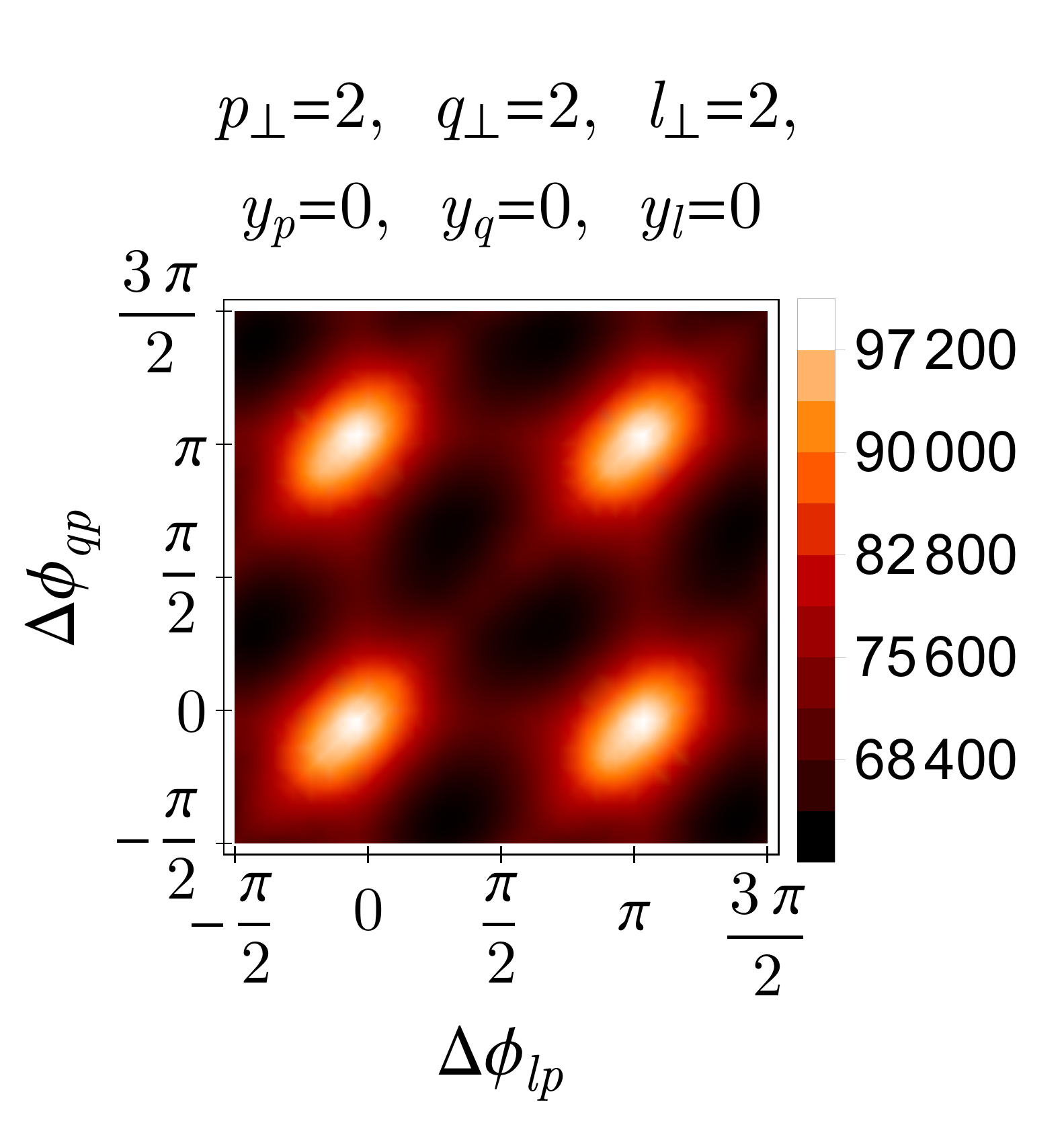} \hspace{-3mm}
\includegraphics[scale=0.325]{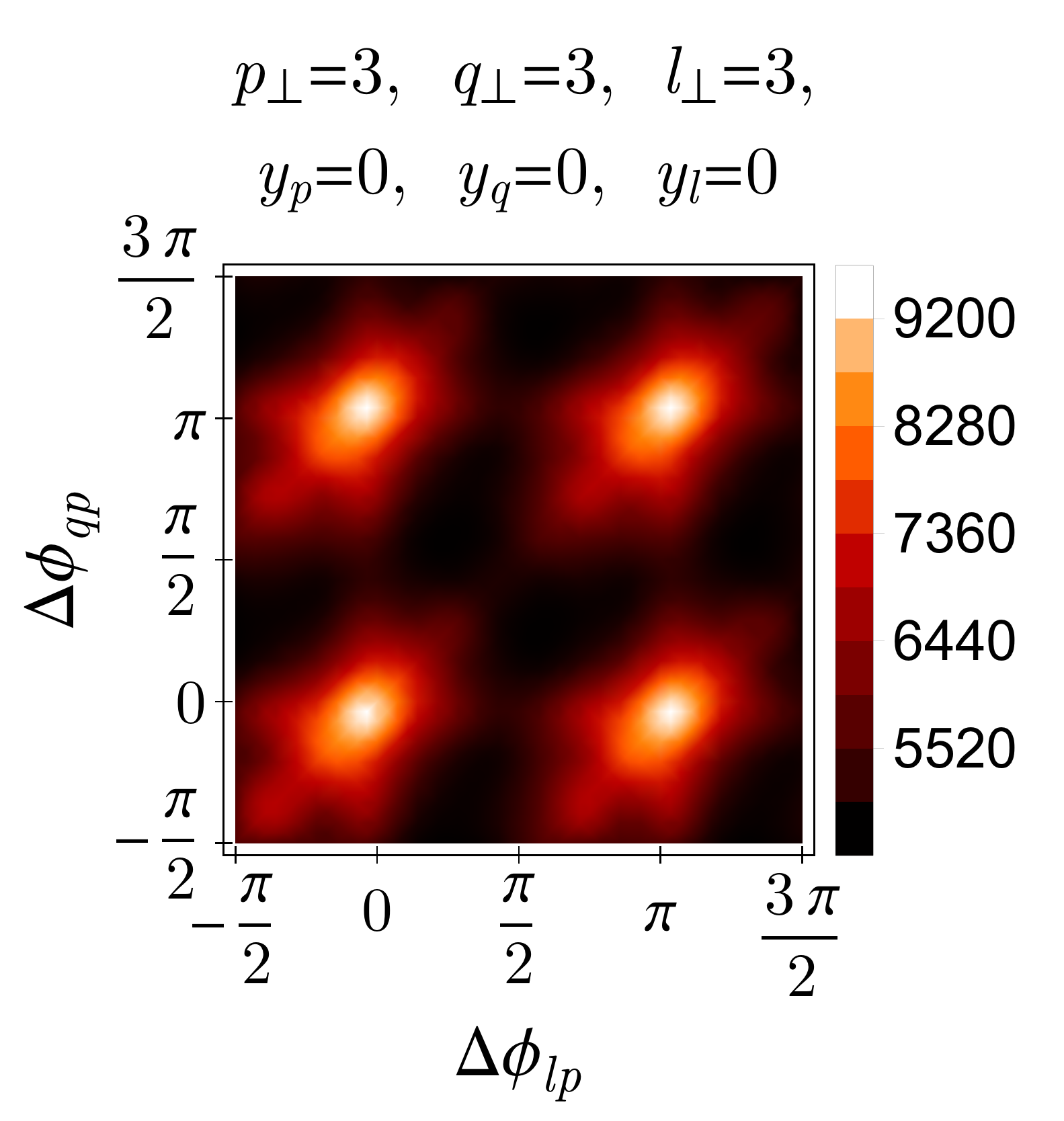} \hspace{-3mm}
\includegraphics[scale=0.325]{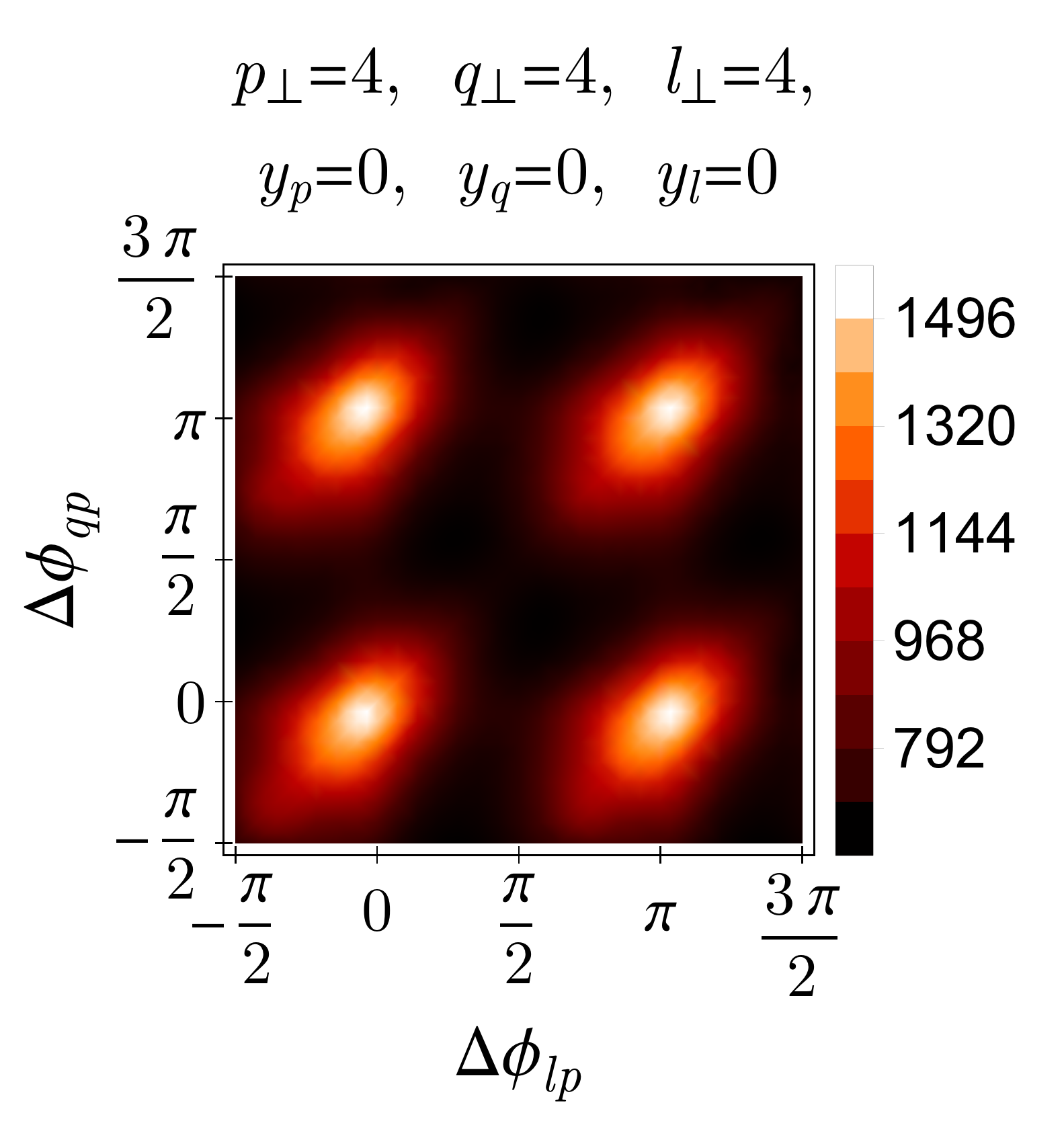}
\par\end{centering}
\caption{(Color online) Triple-gluon azimuthal correlation $C_3(\Delta \phi_{qp},\Delta \phi_{lp})$ for p--Pb collisions at $\sqrt{s}=5.02\tev$ in units of ${\as^3  N_c^3 S_\perp / \pi^{12}  (N_c^2-1)^5}$ [see Eq.~(\ref{C_3})]. The transverse momenta labels are in units of GeV. As the transverse momenta of the gluons increase, the correlation decreases.}
\label{fig:pPbSet1}
\end{figure}

Triple-gluon azimuthal correlations for p--Pb collisions at $\sqrt{s}=5.02\tev$ are shown in Figs.~\ref{fig:pPbSet2} and~\ref{fig:pPbSet1}. While making these two sets of plots we used a UGD that is evolved from the initial scale $Q_0^2=0.168\gev^2$
for the proton, and a UGD that is evolved from the initial scale $Q_0^2=12\times0.168\gev^2$ for the lead nucleus. It can be seen from Figs.~\ref{fig:ppSet1} and~\ref{fig:pPbSet2} that the correlation is stronger in the p--Pb case.
In Fig.~\ref{fig:pPbSet1} we show the triple-gluon azimuthal correlation function for gluons at the same rapidity but with varying the transverse momentum windows. The correlation decreases with increasing transverse momenta, which can also be verified from Eq.~(\ref{C_3}).

%% Fig. 10

\begin{figure}[h!]
\begin{centering}
\includegraphics[scale=0.35]{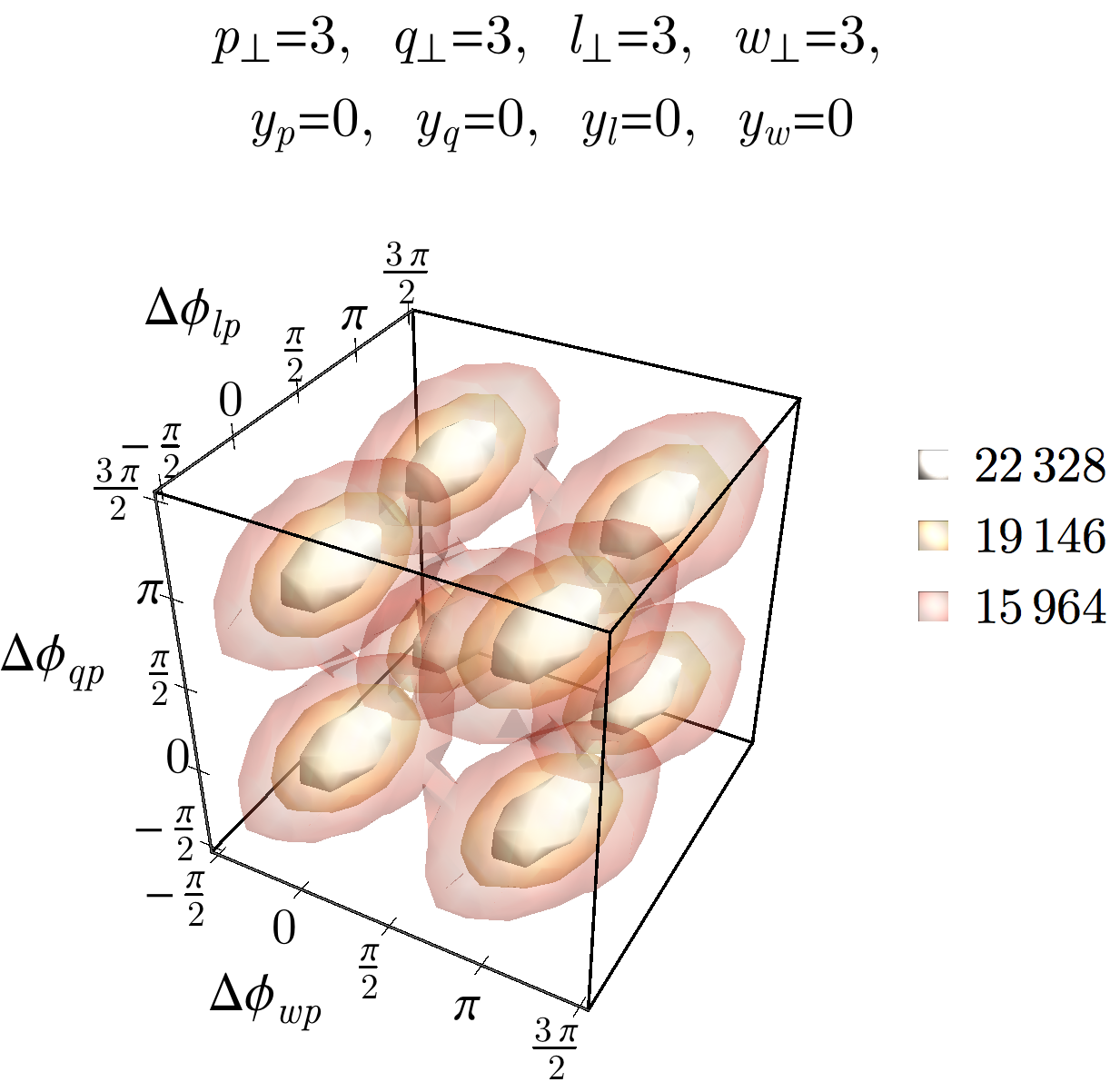} \hspace{2mm}
\includegraphics[scale=0.35]{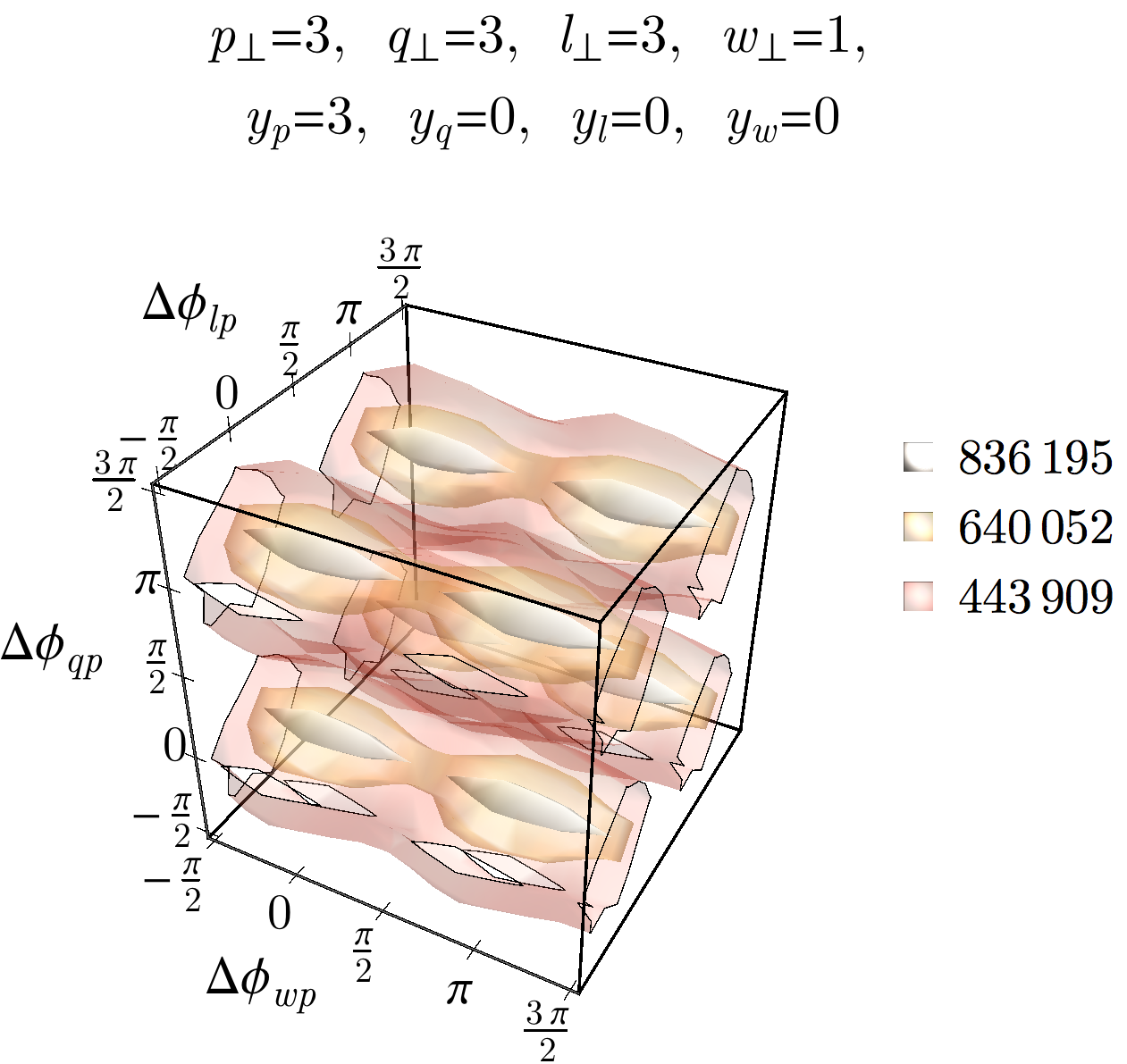}  \hspace{2mm}
\includegraphics[scale=0.35]{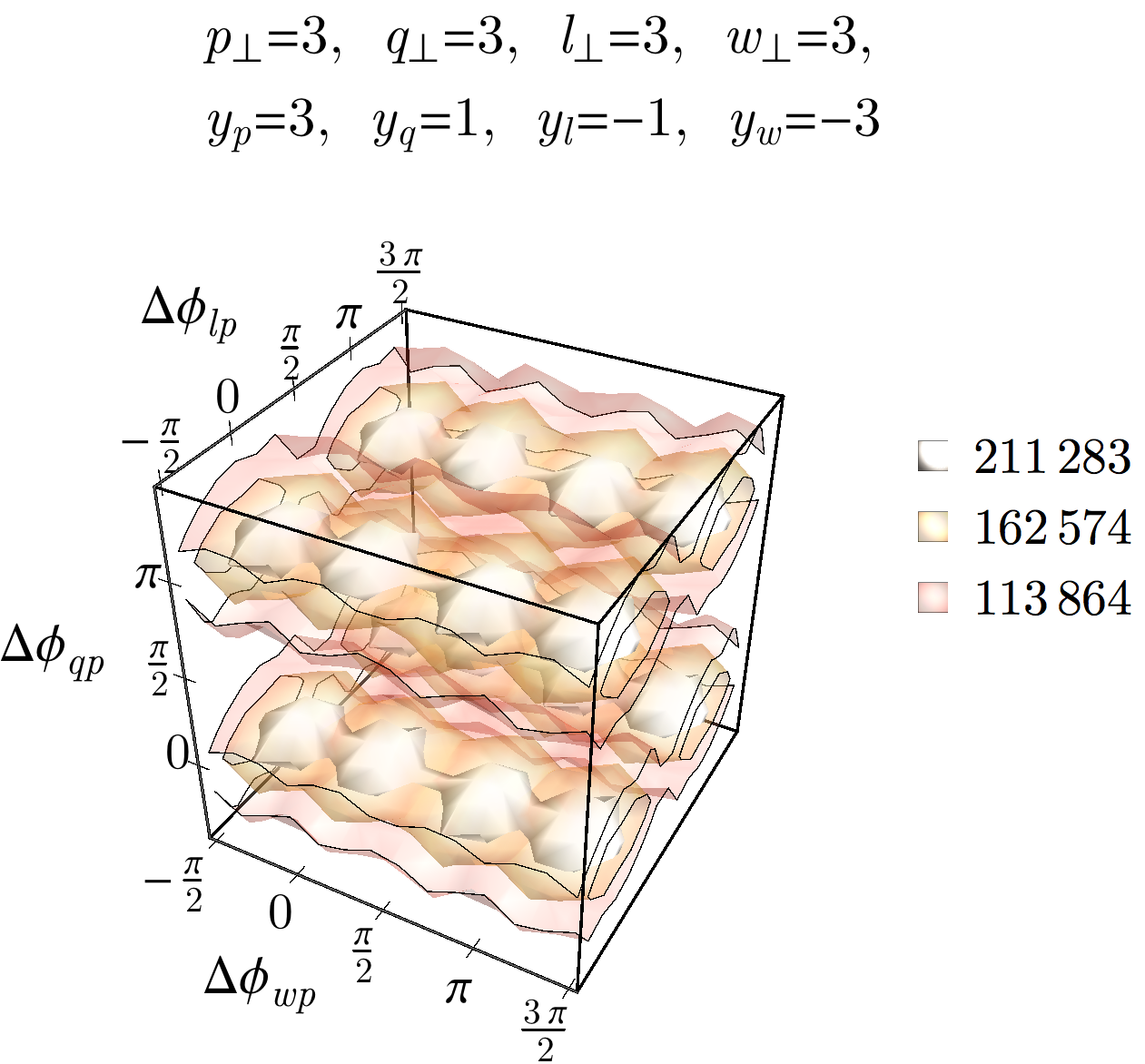}
\par\end{centering}
\caption{(Color online) (left) Quadruple-gluon azimuthal correlation $C_4(\Delta \phi_{qp},\Delta \phi_{lp},\Delta \phi_{wp})$ (left) for p--p collisions at $\sqrt{s}=7\tev$ and (middle and right) for p--Pb collisions at $\sqrt{s}=5.02\tev$. The plots are in units of ${\as^4  N_c^4 S_\perp / \pi^{16}  (N_c^2-1)^7}$ [see Eq.~(\ref{C_4})].
The transverse momenta labels are in units of GeV.}
\label{fig:quadruple-gluon}
\end{figure}

Figure~\ref{fig:quadruple-gluon} shows quadruple-gluon azimuthal correlations for p--p ($\sqrt{s}=7\tev$) and p--Pb ($\sqrt{s}=5.02\tev$) collisions. The three axes in these plots are for the three azimuthal angle differences of the three gluons with respect to the forth gluon with momentum $\pperp$, which is taken to be the trigger. 
%% Fig. 11

%%%%%
%%%%%
%%%%%
\section{Summary and Outlook}
We have calculated the triple- and quadruple-gluon inclusive distributions at arbitrary rapidity and momentum dependence in the gluon saturation regime by using glasma diagrams. These quantities are essential for the calculation of ridges in higher-order particle correlation measurements as well as for the calculation of $v_n$ moments of the final hadron spectra from glasma. We showed our results for the triple- and quadruple-gluon azimuthal correlations for various transverse momenta and rapidities of the produced gluons. 
We predicted that higher dimensional ridges would appear due to gluon saturation in triple- and quadruple-hadron correlations in high-multiplicity p--p and p--Pb collisions at LHC, which have yet to be measured. We left quantitive predictions for a future paper; such a study requires combining glasma correlations due to gluon saturation and  correlations from pQCD in the Regge limit (``mini-jets'' from BFKL or Multi-Regge Kinematics), which is important irregardless of the gluon saturation, and finally convolving the total correlation function with fragmentation functions to obtain the triple- and quadruple-hadron correlations.

\begin{acknowledgements}
I acknowledge the fruitful discussions with Raju Venugopalan and Kevin Dusling. I thank Kevin Dusling for sharing his C/C++ code with me, which helped me while writing my own code in Mathematica. I also thank Michael McNeil Forbes
for the discussions on the color-blind safe, printer and human perception friendly ``gist heat'' colormap that we used in the  two-dimensional density plots.
This work is supported in part by U.S. DOE grant No. DE-FG02-00ER41132.
\end{acknowledgements}

%%%%%%%%% APPENDICES %%%%%%%%%%%
\appendix

%%%%%
%%%%%
%%%%%
\section{Matrix representation of Glasma Diagrams}
\label{appxDiagrams}
Calculation of the relevant glasma diagrams for the triple- and quadruple-gluon correlations is beyond the limit of pen and paper.
The notion of ``adjacency matrices'' from graph theory and its implementation in the computer environment helped us reduce human labor in this work. Any desired diagram can be found among the plethora of topologically distinct glasma diagrams at a given order (double-, triple-, quadruple-gluon etc.) with the help of the adjacency matrix that the diagram is associated with. These matrices tell us if the corresponding diagrams are connected or disconnected, and help us plot any glasma diagram without any extra effort. 
First, the general matrix containing the color and momentum indices for all possible contractions should be  constructed. Then, the color, transverse momentum, rapidity, and charge density structure of a correlation function  can be readily obtained for any diagram. 

In the double-gluon case, for example, the procedure for finding the contributions from Eqs.~(\ref{corr1again}-\ref{rapidity-structure}) to a particular contraction of the color charge densities in Eq.~(\ref{double-gluon-contractions}) is as follows. One has to multiple the adjacency matrix corresponding to a particular diagram with the general matrices for the double-gluon case that are given as 
\be
\bordermatrix{ & p & q & q^* & p^* \cr
                 p &  0  & \delta^2(\Kperp{1} +\Kperp{3})  & \delta^2(\Kperp{1}-\Kperp{4})  &  \delta^2(\Kperp{1}-\Kperp{2})  \cr
                 q &\delta^2(\pperp-\Kperp{1}+\qperp-\Kperp{3}) & 0 &\delta^2(\Kperp{3}-\Kperp{4}) & \delta^2(\Kperp{3}-\Kperp{2})\cr
              q^* & \delta^2(\pperp-\Kperp{1}-\qperp+\Kperp{4}) & \delta^2(\Kperp{4}-\Kperp{3})& 0 & \delta^2(\Kperp{4}+\Kperp{2}) \cr	
              p^* & \delta^2(\Kperp{2}-\Kperp{1}) &\delta^2(\qperp-\Kperp{3}-\pperp+\Kperp{2}) & \delta^2(\pperp-\Kperp{2}+\qperp-\Kperp{4}) & 0}, \label{general-matrix1}
\ee
and

\be
\bordermatrix{ & p & q & q^* & p^* \cr
                 p &  0  & \delta^{bd} \mu^2_{A_1}(y_p,\Kperp{1})  & \delta^{bh} \mu^2_{A_1}(y_p,\Kperp{1})   & \delta^{bf} \mu^2_{A_1}(y_p,\Kperp{1})  \cr
                 q &  \delta^{ec} \mu^2_{A_2}(y_q,\pperp - \Kperp{1})  & 0 & \delta^{dh} \mu^2_{A_1}(y_p,\Kperp{3})  & \delta^{df} \mu^2_{A_1}(y_p,\Kperp{3})  \cr
              q^* & \delta^{ic} \mu^2_{A_2}(y_q,\pperp - \Kperp{1})  & \delta^{ie} \mu^2_{A_2}(y_q,\qperp-\Kperp{3})  & 0 & \delta^{hf} \mu^2_{A_1}(y_p,\Kperp{2})  \cr	
              p^* & \delta^{gc} \mu^2_{A_2}(y_p,\pperp - \Kperp{1})  & \delta^{ge}\mu^2_{A_2}(y_q,\qperp - \Kperp{3}) & \delta^{gi}  \mu^2_{A_2}(y_q,\pperp - \Kperp{2}) & 0}. \label{general-matrix2}
\ee
These matrices are built for the specific color indices and momentum variables of the expression in Eq.~(\ref{double-gluon-contractions})\footnote{The rapidity structure of the triple- and quadruple-gluon matrices is more complex than that of the double-gluon case. Hence, for higher order diagrams one should be more careful about the rapidity indices while constructing the general matrix [see the discussion below Eq.~(\ref{connected})].}. The parts of the matrices above the zeros are for the projectile (nucleus 1) whereas the parts below the zeros are for the target (nucleus 2). The starred momentum 
labels are associated with the starred charge densities in Eq.~(\ref{double-gluon-contractions}).
As an example, multiplying the general matrices in Eqs.~(\ref{general-matrix1}) and (\ref{general-matrix2}) with the adjacency matrices 
\be
\bordermatrix{ & p & q & q^* & p^* \cr
                 p &  0  & 0  & 0   & 1  \cr
                 q &  0  & 0 & 1  & 0  \cr
              q^* & 1  & 0  & 0 & 0  \cr	
              p^* & 0  & 1 & 0 & 0}
\qquad \text{and} \qquad
\bordermatrix{ & p & q & q^* & p^* \cr
                 p &  0  & 0  & 0   & 1  \cr
                 q &  0  & 0 & 1  & 0  \cr
              q^* & 0  & 1  & 0 & 0  \cr	
              p^* & 1  & 0 & 0 & 0} 
\ee
gives the contributions for the connected and disconnected diagrams
shown in Fig.~\ref{fig:double}.

\begin{figure}[t]
\includegraphics[scale=0.65]{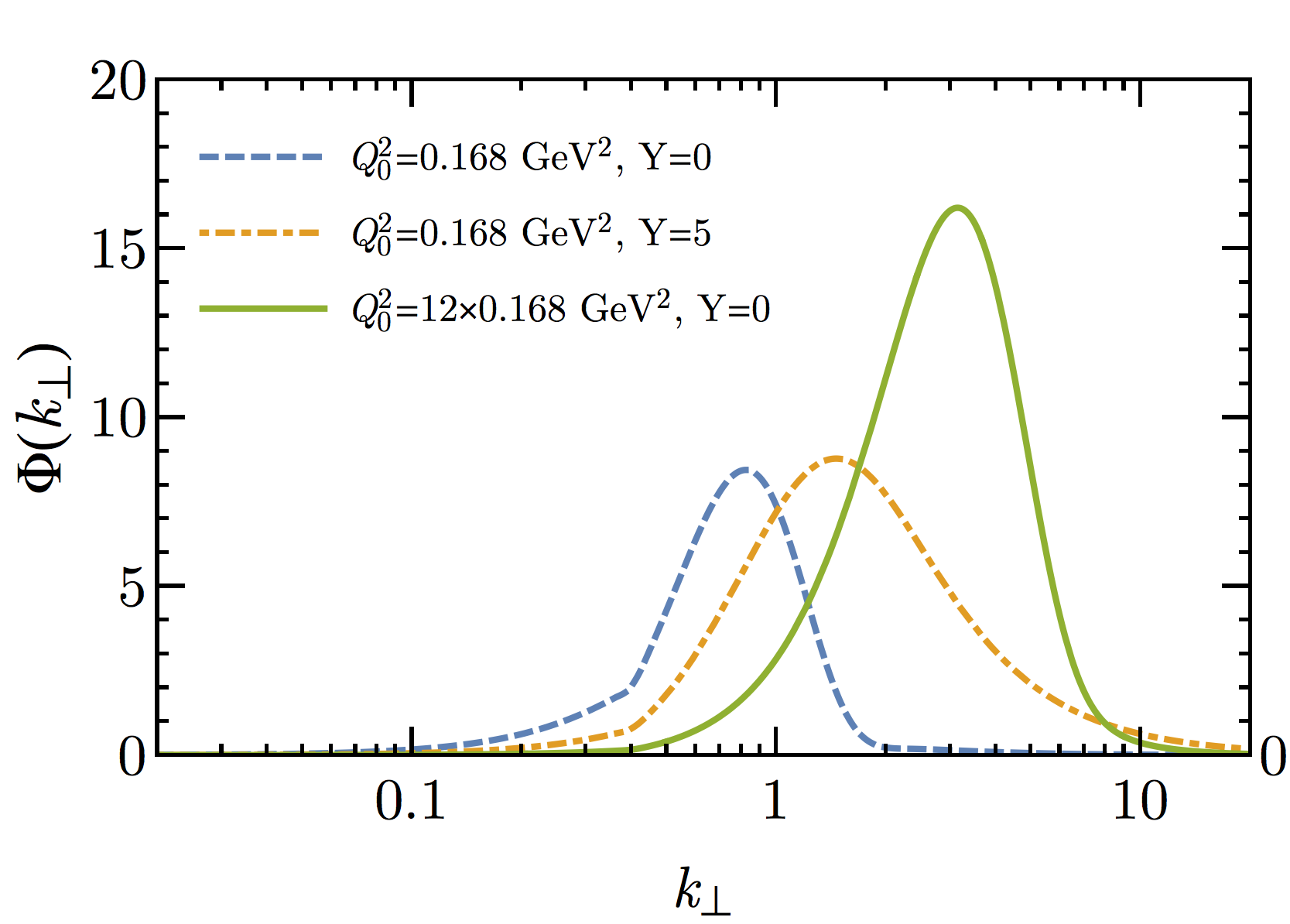}
\caption{Plots of some rcBK unintegrated gluon distributions (UGD). Initial scale of the evolution $Q^2_0$ and parton longitudinal momentum fraction $x$ determine the saturation scale, which is where the UGDs peak. The parton rapidity $Y$ and $x$ are related via $Y=\log(x_0/x)$ where $x_0=0.01$. In this work, we use $Q^2_0=0.168\gev^2$ for the proton and $Q^2_0=12\times0.168\gev^2$ for the lead nucleus. The correlation between $Q^2_0$ and the experimentally relevant quantity $N_\text{trk}^\text{offline}$ has been investigated in Ref. \cite{Dusling:2012wy}.
}
\label{fig:UGDs}
\end{figure}

%%%%%
%%%%%
%%%%%
\section{Glasma Prefactors in the Literature}
\label{appxPrefactors}
The factors of $1/2$ in Eq.~(3.2) in Ref.~\cite{Dusling:2009ni} and in Eq.~(9) in Ref.~\cite{Dumitru:2008wn} are incorrect, and they are corrected later in papers such as \cite{Gelis:2009wh,Dusling:2009ar,Dusling:2009ni}.

Equation~(3.8) for $\langle dN_1/d^2 \pperp dy_p \rangle$ in Ref.~\cite{Dusling:2009ni} is correct but the connected double-gluon correlation function $C(\bo{p},\bo{q})$ given in Eq.~(3.17) in the same paper and Eq.~(3) in Ref.~\cite{Dumitru:2010iy} have an extra factor of $1/4$. This is corrected in Refs.~\cite{Dusling:2012iga,Dusling:2012wy,Dusling:2012cg,Dusling:2012wy,Dusling:2013oia,Dusling:2013oia}, together with the definition $\int_{\kperp} \equiv \int d \kperp$. Those results match with ours as given in Eq.~(\ref{C_2}) in this paper.

%%%%%
%%%%%
%%%%%
\section{rcBK Unintegrated Gluon Distributions}
\label{appxUGD}

In Fig.~\ref{fig:UGDs}, we show some rcBK UGDs evolved from two different initial scales $Q^2_0$; the saturation scale depends on this initial scale. In this work we use the UGDs whose properties have been explained at length in Refs. \cite{Dusling:2012wy,Dusling:2009ni}.

%\raggedright
\bibliographystyle{unsrt}
\bibliography{triple-quadruple-gluon}

\end{document}